\documentclass[referee,sn-mathphys,Numbered]{sn-jnl}% Math and Physical Sciences Reference Style
\usepackage{graphicx}%
\usepackage{multirow}%
\usepackage{amsmath,amssymb,amsfonts}%
\usepackage{amsthm}%
\usepackage{mathrsfs}%
\usepackage[title]{appendix}%
\usepackage{xcolor}%
\usepackage{textcomp}%
\usepackage{manyfoot}%
\usepackage{booktabs}%
\usepackage{algorithm}%
\usepackage{algorithmicx}%
\usepackage{algpseudocode}%
\usepackage{listings}%
\begin{document}

\title[Article Title]{Semiclassical approximation of the Wigner function for the canonical ensemble}


\author*[1]{\fnm{Marcos} \sur{Gil de Oliveira}}\email{marcosgil@id.uff.br}

\author*[2]{\fnm{Alfredo Miguel} \sur{Ozorio de Almeida}}\email{alfredozorio@gmail.com}

\affil[1]{\orgdiv{Departamento de Física}, \orgname{Universidade Federal Fluminense}, \orgaddress{ \city{Niterói}, \postcode{24210-346}, \state{R.J}, \country{Brazil}}}

\affil[2]{\orgname{Centro Brasileiro de Pesquisas Físicas}, \orgaddress{ \city{Rio de Janeiro}, \postcode{22290-180}, \state{R.J.}, \country{Brazil}}}


\abstract{The Weyl-Wigner representation of quantum mechanics allows one to map the density operator in a function in phase space — the Wigner function — which acts like a probability distribution. In the context of statistical mechanics, this mapping makes the transition from the classical to the quantum regimes very clear, because the thermal Wigner function tends to the Boltzmann distribution in the high temperature limit. We approximate this quantum phase space representation of the canonical density operator for general temperatures in terms of classical trajectories, which are obtained  through a Wick rotation of the semiclassical approximation for the Weyl propagator. A numerical scheme which allows us to apply the approximation for a broad class of systems is also developed. The approximation is assessed by testing it against systems with one and two degrees of freedom, which shows that, for a considerable range of parameters, the thermodynamic averages are well reproduced.}




\maketitle

\section{Introduction}

Quantum and classical statistical mechanics differ both in their formulation and in their results. It is not by chance that the first evidence for quantum mechanics is the black-body spectrum derived by Planck \cite{https://doi.org/10.1002/andp.19013090310}. The canonical ensemble, which describes a system in equilibrium with a thermal bath of temperature $T$, is characterized classically through a probability distribution over phase space, the Boltzmann distribution
\begin{equation}
\label{boltzmann distribution}
    P_\beta (\boldsymbol x) = \frac{1}{Z_c}e^{-\beta H_c(\boldsymbol x)},
\end{equation}
where, $\boldsymbol x = (p_1,\ldots,p_d,q_1,\ldots,q_d)$ is a point in the phase space spanned by the coordinates $q_j$ and the momenta $p_j$, $H_c$ is the classical Hamiltonian of the system, $Z_c$ is the classical partition function and $\beta = 1/kT$, $k$ being the Boltzmann's constant. The \textit{quantum} canonical ensemble, on the order hand, is described by the thermal density operator
\begin{equation}
\label{thermal density operator}
    \hat{\rho}_\beta = \frac{1}{Z}e^{-\beta \hat{H}}
\end{equation}
where $\hat{H}$ is the Hamiltonian operator and $Z$ is the quantum partition function. Both \eqref{boltzmann distribution} and \eqref{thermal density operator} allow one to calculate thermodynamic averages, and although the results agree for high temperatures, there is a considerable discrepancy for low ones. With the introduction, by Wigner, of his eponymous function \cite{PhysRev.40.749}, the differences between these two formulations diminished, as it allows one to map the thermal density operator in a function over phase space that works as if it were a probability distribution, though it strongly deviates from \eqref{boltzmann distribution} in the low temperature regime. This proposal further evolved to give a complete formulation of quantum mechanics in phase space \cite{GROENEWOLD1946405,moyal1949quantum}, which we call the Weyl-Wigner representation.  The high temperature limit of the resulting semiclassical approximation of the thermal Wigner function, coincides with the classical distribution \eqref{boltzmann distribution}. 

A further advantage of the Weyl-Wigner formalism is that common observables with
classical correspondence are directly represented by the classical phase space function, or a function that is semiclassically close to it. Thus, there is no
limitation to Hamiltonians with a quadratic momentum dependence: any real phase space function will do.
Moreover, the expectation of the observable is evaluated by a phase space integral, identical to its classical counterpart, except that the Liouville distribution is replaced by the Wigner function. In contrast, the phase space reflection operator, 
\begin{equation}
 \hat{R}_{\boldsymbol x} =
 \int {\rm d}^{N}\boldsymbol \xi_{\boldsymbol q} ~|\boldsymbol q
+ \boldsymbol \xi_{\boldsymbol q}\rangle\langle \boldsymbol q
- \boldsymbol \xi_{\boldsymbol q}| ~
\exp\left[-\frac{2i}{\hbar}
\boldsymbol \xi_{\boldsymbol p} \cdot {\boldsymbol q}\right]  ~.
\end{equation}  
which corresponds
classically to the canonical reflection through a phase space point, is also a quantum observable. Indeed, this displaced parity operator has real eigenvalues
$\pm 1$, which makes it as quantum an observable as a spin. The essential role
that this operator plays in the Wigner-Weyl representation, uncovered by Grossmann \cite{grossmann1976parity} and Royer \cite{royer1977wigner} identifies its expectation with the Wigner function itself:
\begin{equation}
W(\boldsymbol x) \equiv \frac{1}{(2\pi\hbar)^N}~ 
{\rm tr}~\hat{\rho}~\hat{R}_{\boldsymbol x} ~.
\label{Wigfn}    
\end{equation}
In short, the value of the Wigner function at every point in phase space
supplies the expectation of the reflection operator for that point, which is exactly how it has been verified experimentally \cite{bertet2002direct}, by counting even and odd outcomes of phase space reflections on identically prepared states.         

In this paper, we will explore the fact that, by evaluating a propagator $\hat{U}_t = e^{-i t \hat{H}/ \hbar}$ at an imaginary time $-i\theta$, where $\theta = \beta \hbar$ is the \textit{thermal time}, we obtain the operator $\hat{U}_{-i\theta} = e^{-\beta \hat{H}}$, which is proportional to the thermal density operator \eqref{thermal density operator}. This is the so called Wick rotation \cite{ingold2002path,greiner2013field}. We will employ this relation, together with a semiclassical approximation for the propagator, which expresses it in terms of \textit{classical} trajectories, to obtain a semiclassical approximation for the canonical ensemble. In principle, it provides a powerful method for evaluating the thermal density operator at lower temperatures, even for many degrees of freedom, because classical trajectories are computed in parallel. The present initial exploration is limited to two degrees of freedom.

The complexification of the Hamiltonian to adapt it to a thermal, rather than a real evolution is already well established in semiclassical calculations, mainly within the chemical literature \cite{yan2019semiclassical,zhao2002semiclassical,shao2006new,liu2006using,pollak2007new,liu2011approach,liu2011approach2} and  \cite{de1999semiclassical,de2000semiclassical}. Even though the various alternative propagators are also supported by trajectories in phase space,
the end result is the position density matrix. Then a comparison with the classical distribution depends on Wigner's symmetrized Fourier transform over phase space. Furthermore, the complexification is confined to the momentum, which restricts the Hamiltonian to be the sum of a quadratic kinetic term with a potential energy, which excludes even a simple magnetic field.
%The examples given were also restricted to systems whose classical %trajectories can be found analytically. (creio que não vale para as novas refs).
In contrast, the complexification employed here is, of need, much less simple (even in the case of the quadratic momentum dependence favoured by the position representation), so as to accommodate arbitrary Hamiltonians, for which the real time evolution is inaccessible to the differential Schr\"odinger equation.
Thus, together with computational tests for standard Hamiltonians, we test the thermal averages of Birkhoff normal forms \cite{arnol2013dynamical,de1988hamiltonian}, which include the quartic Kerr Hamiltonian \cite{haroche2006exploring}; in its turn, the unit cell of the many-body Bose-Hubbard Hamiltonian \cite{schlagheck2019enhancement}. 

%beyond the fact that it makes the connection with classical statistical %mechanics very straightforward. A discussion of the canonical ensemble in this %representation can also be seen \cite{nicacio2021weyl}, but again with a %restriction — only quadratic hamiltonians are considered. 

This paper is a followup on \cite{OZORIODEALMEIDA2021132951}, where the core results of our current approach were first proposed. Here, we bridge the gaps that remained, which then allows us to devise a computational scheme that opens the possibility of applying our approximation to a vast number of cases. These new developments were achieved during a master's degree, and first appeared on the thesis \cite{gil2023master}.

The presentation is then structured as follows: in section \ref{The Weyl-Wigner representation and semiclassical approximations} we discuss elements of the Weyl-Wigner representation and introduce a semiclassical approximation for the propagator. In section \ref{The canonical ensemble in phase space} we particularize this discussion for the canonical ensemble, and show how the approximation for the propagator generates an approximation for the thermal density operator through the Wick rotation. In section \ref{Normal Forms}, we apply our approximations for normal forms, which are a class of systems for which one has explicit expressions for the required quantities. In section \ref{Double Phase Space} we reformulate the calculation of the trajectories in terms of a duplicated phase space, which is more amenable to a computational treatment, and develop a complete numerical method that allows us, in principle, to apply our approximation for systems with an arbitrary hamiltonian. In sections \ref{Morse System} and \ref{Nelson System}, we use this numerical scheme to apply the approximation to the Morse system, which has one degree of freedom, and to the Nelson system, which has two.

\section{The Weyl-Wigner representation and semiclassical approximations}
\label{The Weyl-Wigner representation and semiclassical approximations}

The Weyl-Wigner representation of quantum mechanics is based on the reflection operators
\begin{equation}
    \hat{R}_{\boldsymbol x} = \int \frac{d\boldsymbol{\xi}}{(4\pi \hbar)^d} \exp \left[ \frac{i}{\hbar} \boldsymbol{\xi} \wedge \left( \hat{\boldsymbol x}- \boldsymbol x \right) \right],
\end{equation}
which corresponds classically to the transformation $R_{\boldsymbol x} : \boldsymbol{x}_- \mapsto 2\boldsymbol x-\boldsymbol{x}_-$. Here, $\hat{\boldsymbol x} = (\hat{p}_1,\ldots,\hat{p}_d,\hat{q}_1,\ldots,\hat{q}_d)$ is a vector formed by the position and momentum operators and $\wedge$ denotes the \textit{wedge product}, defined by $\boldsymbol \xi \wedge \boldsymbol x = \left(\boldsymbol{J \xi} \right) \cdot \boldsymbol x$, with
\begin{equation}
    \boldsymbol{J} = \left( 
    \begin{array}{c|c} 
        \boldsymbol{0} & -\boldsymbol{I}_d \\ 
        \hline 
        \boldsymbol{I}_d & \boldsymbol{0} 
    \end{array} 
    \right),
\end{equation}
where $\boldsymbol{I}_d$ denotes the $d \times d$ identity matrix.

The Wigner symbol $O(\boldsymbol x)$ of an operator $\hat{O}$ is then given by
\begin{equation}
    O(\boldsymbol x) = 2^d \text{Tr } \left( \hat{O} \hat{R}_{\boldsymbol x} \right).
\end{equation}
Furthermore, the Wigner function is a quantity proportional to the Wigner symbol of the density operator
\begin{equation}
\label{definition wigner}
    W(\boldsymbol x) = \frac{\rho(\boldsymbol x)}{(2\pi \hbar)^d} = \frac{1}{(\pi \hbar)^d} \text{Tr } \left( \hat{\rho} \hat{R}_{\boldsymbol x} \right)
\end{equation}
and can be used to calculate quantum averages
\begin{equation}
\label{valor esperado wigner}
    \left< \hat{O} \right> = \text{Tr } \left(\hat{\rho}\ \hat{O}\right) =  \int d\boldsymbol{x} W\left(\boldsymbol{x}\right) O\left(\boldsymbol{x}\right)
\end{equation}
as if it were a probability distribution.

As an example, we observe that the Wigner function for the eigenstates of the harmonic oscillator, defined by a hamiltonian $\hat{H} = \omega \left( \hat{p}^2+\hat{q}^2 \right)/2$, are given by
\begin{equation}
\label{função de wigner do oscilador harmônico}
    W_n(\boldsymbol{x}) = \frac{(-1)^n}{\pi \hbar} e^{-\boldsymbol{x}^2/\hbar}L_n\left( \frac{2 \boldsymbol{x}^2}{\hbar} \right),
\end{equation}
where $L_n$ is the $n$th Laguerre polynomial \cite{GROENEWOLD1946405}. 

A striking feature of the Wigner representation is the fact that the Wigner symbol of operators of the form $f\left( \hat{\boldsymbol{p}} \right) + g\left( \hat{\boldsymbol{q}} \right)$ is simply $f\left( \boldsymbol{p} \right) + g\left( \boldsymbol{q} \right)$, which is exactly the corresponding classical variable. This is not a general result, as there can be corrections in form of power series of $\hbar$. A useful formula for calculating more complicated Wigner symbols is the Groenewold rule \cite{de1998weyl}
\begin{equation}
\label{groenewold}
    \begin{aligned}
        O_2 \cdot O_1 \left(\boldsymbol{x}\right) &= O_2\left(\boldsymbol{x}+\frac{i \hbar}{2}\boldsymbol{J}\frac{\partial}{\partial \boldsymbol{x}}\right)  O_1\left(\boldsymbol{x}\right) \\ & =O_1\left(\boldsymbol{x}-\frac{i \hbar}{2}\boldsymbol{J}\frac{\partial}{\partial \boldsymbol{x}}\right)  O_2\left(\boldsymbol{x}\right).
    \end{aligned}
\end{equation}

The Wigner symbol $U_t(\boldsymbol x)$ of the propagator
\begin{equation}
\label{propagator}
    \hat{U}_t = e^{-i t \hat{H}/ \hbar}
\end{equation}
is called the Weyl propagator, and, for short enough times, has the semiclassical approximation \cite{de1998weyl}
\begin{equation}
\label{sc weyl propagator}
    U_t(\boldsymbol x)_{SC} = \left| \det \left( \boldsymbol{I}_{2d} \pm \boldsymbol{J} \boldsymbol{B}_t   \right)\right|^{1/2} \exp \left[ \frac{i}{\hbar} S_t(\boldsymbol x) \right].
\end{equation}
Here, $S_t(\boldsymbol x) = S(\boldsymbol x,t)$ is the so called (centre) action, and
\begin{equation}
    \boldsymbol{B}_t = \frac{1}{2} \frac{\partial^2 S_t}{\partial \boldsymbol x^2}
\end{equation}
is proportional to its hessian. The action has the role of a generating function for the classical hamiltonian flow. It indirectly specifies the transformation $\boldsymbol x_- \mapsto \boldsymbol x_+$ by giving the chord
\begin{equation}
    \label{definition chord}
    \boldsymbol \xi = \boldsymbol x_+ - \boldsymbol x_-
\end{equation}
in terms of the centre
\begin{equation}
    \label{definition centre}
    \boldsymbol x = \frac{\boldsymbol x_- + \boldsymbol x_+}{2}
\end{equation}
through the relation
\begin{equation}
    \label{chord in terms of centre}
    \boldsymbol \xi(\boldsymbol x,t) = - \boldsymbol J \frac{\partial S_t}{\partial \boldsymbol x}.
\end{equation}

As going back in time simply reverses the hamiltonian flow, we must have $\boldsymbol \xi(\boldsymbol x,t) = -\boldsymbol \xi(\boldsymbol x,-t)$, from which we conclude that the centre action must be an odd function of $t$. Furthermore, Hamilton's equations
\begin{equation}
    \dot{\boldsymbol x} = \boldsymbol J \frac{\partial H}{\partial \boldsymbol x}
\end{equation}
imply that, for short times $t$, we have
\begin{equation}
    \boldsymbol \xi \approx t \boldsymbol J \frac{\partial H}{\partial \boldsymbol x},
\end{equation}
from which we get
\begin{equation}
\label{short time approximation}
    S(\boldsymbol x,t) = -tH(\boldsymbol x) + \mathcal{O} \left( t^3 \right).
\end{equation}
In general, the action is given by
\begin{equation}
    S(\boldsymbol x,t) = \Delta(\boldsymbol x,t) - Et
\end{equation}
where $\Delta$ is the symplectic area of the region between the trajectory and the chord and $E$ is the energy of the trajectory.

As an example for quadratic hamiltonians
\begin{equation}
     H(\boldsymbol{x}) = \frac{1}{2} \boldsymbol{x} \cdot \boldsymbol{\mathcal{H}}_0 \boldsymbol{x}
\end{equation}
the action is also quadratic and given by
\begin{equation}
    S_t(\boldsymbol x) = \boldsymbol{x} \cdot \boldsymbol B_t \boldsymbol{x}; \ \ \ \boldsymbol B_t = \boldsymbol{J}\tanh \left(\frac{t}{2}\boldsymbol{J} \boldsymbol{\mathcal{H}}_{\boldsymbol{0}}\right).
\end{equation}

It is important to observe that, for this class of systems, the semiclassical approximations are actually exact \cite{de1998weyl}. For a harmonic oscillator with frequency $\omega$, we have $\boldsymbol{\mathcal{H}}_0 = \omega \boldsymbol{I}_{2d}$, and we obtain an exact Weyl propagator
\begin{equation}
\label{weyl propagator ho}
    U_t(\boldsymbol x) = \sec \left( \omega t /2 \right) \exp \left[ -\frac{i}{\hbar} \tan\left( \omega t /2 \right) \boldsymbol x^2  \right].
\end{equation}
We see that, when $\omega t \to (2n+1)\pi$, we have
\begin{equation}
    \left| \det \left( \boldsymbol{I}_{2d} \pm \boldsymbol{J} \boldsymbol{B}_t  \right)\right|^{1/2} = \sec \left( \omega t /2 \right) \to \infty.
\end{equation}
The set of points where this divergence occurs is called a caustic, and it signals the breakdown of the description of the canonical transformation by the centre generating function. For the harmonic oscillator, the hamiltonian flow in these instants is simply a reflection $\boldsymbol x_- \mapsto \boldsymbol x_+ = - \boldsymbol x_-$, and therefore, for every pair $(\boldsymbol x_-,\boldsymbol x_+)$ we get the same centre $\boldsymbol x = \boldsymbol 0$. In this case, then, the caustics are the entire phase space, but, for non quadratic hamiltonians, these divergences may be restricted to a lower dimensional sub-manifold. In general, after crossing a caustic, there may be more than one chord for each centre, and the approximation for the propagator becomes a sum of terms like \eqref{sc weyl propagator}, where one must include an extra \textit{Maslov phase} in the exponents \cite{de2014metaplectic,nicacio2017unified}.

It is possible to show \cite{de1998weyl} that the jacobian 
\begin{equation}
    \boldsymbol M_t = \frac{\partial \boldsymbol{x}_+}{\partial \boldsymbol x_-}
\end{equation}
 of the hamiltoninan flow, which is a symplectic matrix \cite{arnol2013mathematical}, is related to $\boldsymbol B_t$ through the Cayley parametrization \cite{arnol2013dynamical}
 \begin{equation}
     \boldsymbol M_t = \frac{\boldsymbol{I}_{2d} - \boldsymbol J \boldsymbol B_t}{\boldsymbol{I}_{2d} + \boldsymbol J \boldsymbol B_t},
 \end{equation}
 allowing us to rewrite \eqref{sc weyl propagator} as 
 \begin{equation}
 \label{sc weyl propagator in terms of M}
     U_t(\boldsymbol x)_{SC} = \frac{2^d}{\left| \det \left( \boldsymbol{I}_{2d} + \boldsymbol M_t   \right)\right|^{1/2}} \exp \left[ \frac{i}{\hbar} S_t(\boldsymbol x) \right],
 \end{equation}
 which will be the most convenient form of the semiclassical propagator to work with.

\section{The canonical ensemble in phase space}
\label{The canonical ensemble in phase space}

As mentioned in the introduction, this work is based on the evaluation of the semiclassical approximation for the propagator \eqref{sc weyl propagator in terms of M}, at the imaginary time $t = -i\theta$, $\theta = \hbar \beta$ being the thermal time. We then obtain a semiclassical approximation for $e^{-\beta \hat{H}}(\boldsymbol x)$:
 \begin{equation}
     e^{-\beta \hat{H}}(\boldsymbol x)_{SC} = \frac{2^d}{\left| \det \left( \boldsymbol{I}_{2d} + \boldsymbol M_{-i\theta}   \right)\right|^{1/2}} \exp \left[ \frac{1}{\hbar} S_\theta^E(\boldsymbol x) \right],
 \end{equation}
where we have defined the euclidean action $S^{E}_{\theta} = i S_{-i\theta}$, which is necessarily real, as $S$ is an odd function of $t$. For the harmonic oscillator, we get, using \eqref{weyl propagator ho},
\begin{equation}
    e^{-\beta \hat{H}}(\boldsymbol x) = \text{sech} \left( \omega \theta /2 \right) \exp \left[ -\frac{1}{\hbar} \text{tanh}\left( \omega \theta /2 \right) \boldsymbol x^2  \right].
\end{equation}

It is interesting to note that, by using the the short time approximation \eqref{short time approximation} and setting $\boldsymbol M_t \approx \boldsymbol{I}_{2d}$, we get
\begin{equation}
    e^{-\beta \hat{H}}\left( \boldsymbol{x} \right)_{SC} \approx \exp \left[ -\beta H \left(\boldsymbol{x} \right) \right] \approx \exp \left[ -\beta H_c \left(\boldsymbol{x} \right) \right],
\end{equation}
that is, for high temperatures, we recover the \textit{classical} canonical ensemble.

In this framework, the thermodynamic expectation values
\begin{equation}
    \left< \hat{O} \right> = \text{Tr } \left(\hat{\rho}_\beta \hat{O}\right) = \frac{\text{Tr } \left(e^{-\beta \hat{H}} \hat{O}\right)}{\text{Tr } e^{-\beta \hat{H}}},
\end{equation}
are completely determined if one is able to calculate expressions of the form $\text{Tr } \left( U_t \hat{O} \right)$ for imaginary $t$, which has an approximation
\begin{equation}
    \label{trace of U O}
    \text{Tr } \left( \hat{U}_t \hat{O} \right)_{SC} = \frac{1}{(\pi \hbar)^d}\int d \boldsymbol{x} \frac{ \ e^{i S_t(\boldsymbol{x}) / \hbar}  O(\boldsymbol{x})}{\left|\det \left[ \boldsymbol{I} + \boldsymbol{M}_{t}\right] \right|^{1/2} } .
\end{equation}

One of the first problems that appears when dealing with semiclassical approximations, and can already be seen in \eqref{trace of U O}, is the fact that the relevant trajectories are specified by boundary-conditions — in the case of the Wigner representation, we specify the centre $\boldsymbol x$ defined by the endpoints of the trajectory — that can be satisfied by more than one orbit, and are much more difficult to solve than an initial value problem. This is the so called \textit{root search problem}.

\begin{figure}[ht]
    \centering
    \includegraphics[width=0.60\linewidth]{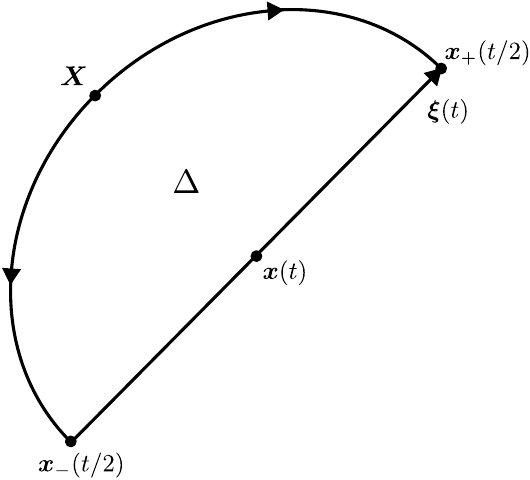}
    \caption{Midpoint representation.}
    \label{midpoint representation}
\end{figure}

There are a few methods that can be used to circumvent this question, including the Initial and Final Value Representations \cite{de2013initial}. Here, we briefly discuss a method that is specially adapted for he calculation of \eqref{trace of U O}, which we call the \textit{midpoint representation}, and consists of a mere change of variables, that we explain in what follows. We start with a point $\boldsymbol X$ in phase space — the midpoint — from which we propagate a trajectory $\boldsymbol x_+(t)$, that evolves forward in time, and a trajectory $\boldsymbol x_-(t)$, which evolves backwards, as illustrated in figure \ref{midpoint representation}. 

In other words, $\boldsymbol x_\pm(t)$ satisfy the pair of initial value problems
\begin{equation}
    \label{x pm definition}
    \dot{\boldsymbol x}_\pm(t) = \pm \boldsymbol{J} \frac{\partial H}{\partial \boldsymbol x}; \ \ \ \boldsymbol x_\pm(0) = \boldsymbol X .
\end{equation}
From this trajectory, we construct a centre
\begin{equation}
    \boldsymbol x(t) = \frac{\boldsymbol x_+(t/2) + \boldsymbol x_-(t/2)}{2}
\end{equation}
and a chord
\begin{equation}
    \boldsymbol \xi(t) = \boldsymbol x_+(t/2) - \boldsymbol x_-(t/2).
\end{equation}

Then, the transformation $\boldsymbol x \mapsto \boldsymbol X$ has a jacobian determinant
\begin{equation}
    \begin{aligned}
        \det \frac{\partial \boldsymbol x}{\partial \boldsymbol X} &= \frac{1}{2^{2d}} \det \left( \boldsymbol M_{t/2} + \boldsymbol M_{-t/2}\right) \\
        &= \frac{1}{2^{2d}} \det \left( \boldsymbol{I}_{2d} + \boldsymbol M_{t}\right),
    \end{aligned}
\end{equation}
where we have used the fact that symplectic matrices, such as $\boldsymbol M$, have unit determinant, and that the composition law $\boldsymbol M_{t_2} \boldsymbol M_{t_1} = \boldsymbol M_{{t_1+t_2}}$ holds. Performing this change of variables in \eqref{trace of U O}, we arrive at
\begin{equation}
    \label{trace of U O in terms of X}
    \begin{aligned}
        \text{Tr } \left( \hat{U}_t \hat{O} \right)_{SC} = \frac{1}{(2\pi \hbar)^d}\int d \boldsymbol{X}& \left| \frac{\partial \boldsymbol x}{\partial \boldsymbol X}\right|^{1/2} e^{i S_t(\boldsymbol{x}) / \hbar}  O(\boldsymbol{x}) .
    \end{aligned}
\end{equation}

Furthermore, in this framework, the area $\Delta$ can be explicitly calculated as \cite{OZORIODEALMEIDA2021132951}
\begin{equation}
    \label{area}
    \Delta\left[\boldsymbol{x}\left( \boldsymbol{X} \right),t\right] = \int_0^{t/2} \boldsymbol{\xi}\left(\boldsymbol{X},t'\right) \wedge \dot{\boldsymbol{x}}\left(\boldsymbol{X},t'\right) dt', 
\end{equation}
from which we obtain the action
\begin{equation}
    S_t\left[\boldsymbol{x}\left( \boldsymbol{X} \right)\right] = \int_0^{t/2} \boldsymbol{\xi}\left(\boldsymbol{X},t'\right) \wedge \dot{\boldsymbol{x}}\left(\boldsymbol{X},t'\right) dt'-tH\left( \boldsymbol{X}\right).
\end{equation}

Beyond the fact that, now, all quantities can be determined by the initial value problem \eqref{x pm definition}, we see that another advantage of this representation is the property that, at caustics, the integrand in \eqref{trace of U O in terms of X} is now zero, instead of infinite, as was the case in expression \eqref{trace of U O}.

\section{Normal Forms}
\label{Normal Forms}

Here, we discuss a class of systems for which we are able to obtain explicit analytical results for the trajectories \eqref{x pm definition}, which allow a direct evaluation of \eqref{trace of U O in terms of X} at imaginary times. 

A one dimensional \textit{classical} hamiltonian written as
\begin{equation}
\label{classical normal form}
    \begin{aligned}
        H(\boldsymbol{x}) &= \omega \left( \frac{p^2+q^2}{2} \right) + H_2 \left( \frac{p^2+q^2}{2} \right)^2 \\ &+ H_3 \left( \frac{p^2+q^2}{2} \right)^3 + \cdots = F\left( \frac{\boldsymbol{x}^2}{2} \right)
    \end{aligned}
\end{equation}
is said to be in Birkhoff normal form \cite{arnol2013dynamical,de1988hamiltonian}. Its orbits are circles in phase space, and, therefore, by a simple geometric argument, one may find explicit expressions for all the ingredients of the semiclassical approximation, as done in \cite{OZORIODEALMEIDA2021132951}. Defining the action variable
\begin{equation}
    J = \frac{\boldsymbol{X}^2}{2}
\end{equation}
and
\begin{equation}
    \omega(J) = F^\prime(J),
\end{equation}
we find the euclidean action
\begin{equation}
\label{euclidean eaction nf}
    S_\theta^E\left[ \boldsymbol{x}\left( \boldsymbol{X}\right) \right] = \left[ \omega \theta - \sinh \left(\omega \theta\right) \right] J - \theta F\left( J \right),
\end{equation}
the centre
\begin{equation}
    \label{thermal centre}
    \boldsymbol{x}\left( \boldsymbol{X},-i\theta/2 \right) = \cosh \left (\frac{\omega \theta}{2} \right) \boldsymbol{X},
\end{equation}
and the jacobian determinant
\begin{equation}
    \begin{aligned}
        &\det \frac{\partial \boldsymbol{x}}{\partial \boldsymbol{X}}(\boldsymbol{X},-i\theta/2) \\ &= \cosh^2 \left( \frac{\omega \theta}{2} \right)\left[1 + J \omega'  \theta \tanh \left( \frac{\omega \theta}{2} \right) \right].
    \end{aligned}
\end{equation}

\textit{Quantum} hamiltonians that can be written as
\begin{equation}
\label{quantum normal form}
    \hat{H} = G\left( \frac{\hat{p}^2+\hat{q}^2}{2} \right)
\end{equation}
have a Wigner symbol that is a Birkhoff's normal form, but, except in the case of the harmonic oscillator, the function $F$ in \eqref{classical normal form} \textit{does not} coincide with $G$. Formulas for the calculation of the symbols are presented in \ref{Símbolo de Wigner para formas normais}.

Some of the quantum properties of this class of systems are also readily obtained — they share their eigenstates with the harmonic oscillator, and for a quantum normal form described by a hamiltonian $G\left[\left( \hat{p}^2 + \hat{q}^2 \right)/2\right]$, the eigenenergies are simply $G\left[\hbar\left( n + 1/2 \right)\right]$. This simplicity makes the comparison of our semiclassical approximation with the quantum result straightforward.

It is instructive to analyze the behaviour of our approximation in the low temperature limit. By squaring \eqref{thermal centre}, we obtain
\begin{equation}
\label{J em função de x}
    \frac{\boldsymbol{x}^2}{2} = J \cosh^2 \left[ \frac{\theta \omega\left(J\right)}{2} \right].
\end{equation}
Interpreting this equation as an implicit definition of $J\left( \boldsymbol{x},\theta \right)$, we see that, for a fixed $\boldsymbol{x}$, in order for the right side of the equality to remain finite, we must have
\begin{equation}
    \lim_{\theta \to \infty} J\left( \boldsymbol{x},\theta \right) = 0 \text{ or } \lim_{\theta \to \infty} \omega \left[ J\left( \boldsymbol{x},\theta \right) \right] = 0.
\end{equation}
Therefore, for systems that satisfy $\omega \left( J \right) \ne 0 \ \forall \ J$, we see that, according to \eqref{J em função de x}, $J$ has the asymptotic behaviour
\begin{equation}
    J \approx \frac{\boldsymbol{x}^2}{2} \text{sech}^2 \left[ \frac{\theta \omega\left(0\right)}{2} \right], \ \ \ \theta \omega\left(0\right) \gg 1
\end{equation}

Substituting this expression in \eqref{euclidean eaction nf} and taking the limit $\theta \to \infty$, we obtain, except for a constant term in $\boldsymbol{x}$,
\begin{equation}
    S_\infty^E\left( \boldsymbol{x}\right) = -\boldsymbol{x}^2
\end{equation}
and the semiclassical approximation converges to the quantum result, that is, the semiclassical approximation for the thermal Wigner function tends to the Wigner function of the ground state of the harmonic oscillator, given in \eqref{função de wigner do oscilador harmônico}. Therefore, we see that, at least for normal forms with  $\omega \ne 0$, the semiclassical approximation is well anchored in both the high temperature limit, as it coincides with the classical result, and in the low temperature one, as it correctly predicts the ground state. It then remains to analyze its behaviour for intermediate temperatures.

\subsection{Kerr system}

The simplest case, beyond the harmonic oscillator, of a system governed by a normal form is probably the Kerr system, whose Hamiltonian is
\begin{equation}
    \hat{H} = \hbar \omega_0 \left[ \left( \frac{\hat{p}^2+\hat{q}^2}{2\hbar} \right) + \chi \left( \frac{\hat{p}^2+\hat{q}^2}{2\hbar} \right)^2 \right],
\end{equation}
where $\chi>0$ is a dimensionless parameter and $\omega_0>0$ is a frequency. This Hamiltonian models the propagation of light through a medium with cubic electric susceptibility \cite{haroche2006exploring}. The time evolution of coherent states under its action is known \cite{PhysRevLett.57.13,AVERBUKH1989449}, and the corresponding Wigner function has been experimentally measured \cite{kirchmair2013observation}. This evolution has also been successfully simulated, in the case with $\chi \to \infty$, utilizing semiclassical techniques \cite{PhysRevA.99.042125}.
A further point of interest is that the Hamiltonian for the Bose-Hubbard chain \cite{schlagheck2019enhancement} in many-body physics can be considered as a coupling of Kerr oscillators, which highlights the importance of exploring semiclassical methods for Hamiltonians with non-quadratic momenta.

We note that, in this case, the Wigner symbol of the Hamiltonian only coincides with its classical counterpart within a constant term, as, with the aid of \eqref{groenewold}, one finds
\begin{equation}
    H(p,q) = \hbar \omega_0 \left[ \left( \frac{p^2 + q^2}{2\hbar} \right) + \chi \left( \frac{p^2 + q^2}{2\hbar} \right)^2 - \frac{\chi}{4} \right].
\end{equation}

Identifying
\begin{equation}
    F(J) = \hbar\omega_0 \left[ \frac{J}{\hbar} + \chi\left(\frac{J}{\hbar} \right)^2 - \frac{\chi}{4} \right],
\end{equation}
we see that
\begin{equation}
    \omega(J) = F'(J) = \omega_0 \left( 1 + \chi \frac{J}{\hbar} \right) \ge \omega_0 > 0
\end{equation}
and, therefore, we should expect a good result for low temperatures. Furthermore, when $\chi \ll 1$, we also expect a good result, as, when $\chi \to 0$, we recover the harmonic oscillator, for which the semiclassical approximation is exact. We also note that, because 
\begin{equation}
    \omega'(J) = \chi \omega_0/\hbar > 0,
\end{equation}
one sees that $\det \partial \boldsymbol{x}/ \partial \boldsymbol{X} >0$, that is, there are no caustics for imaginary time.

\begin{figure}[ht]
    \centering
    \includegraphics[width=.95\linewidth]{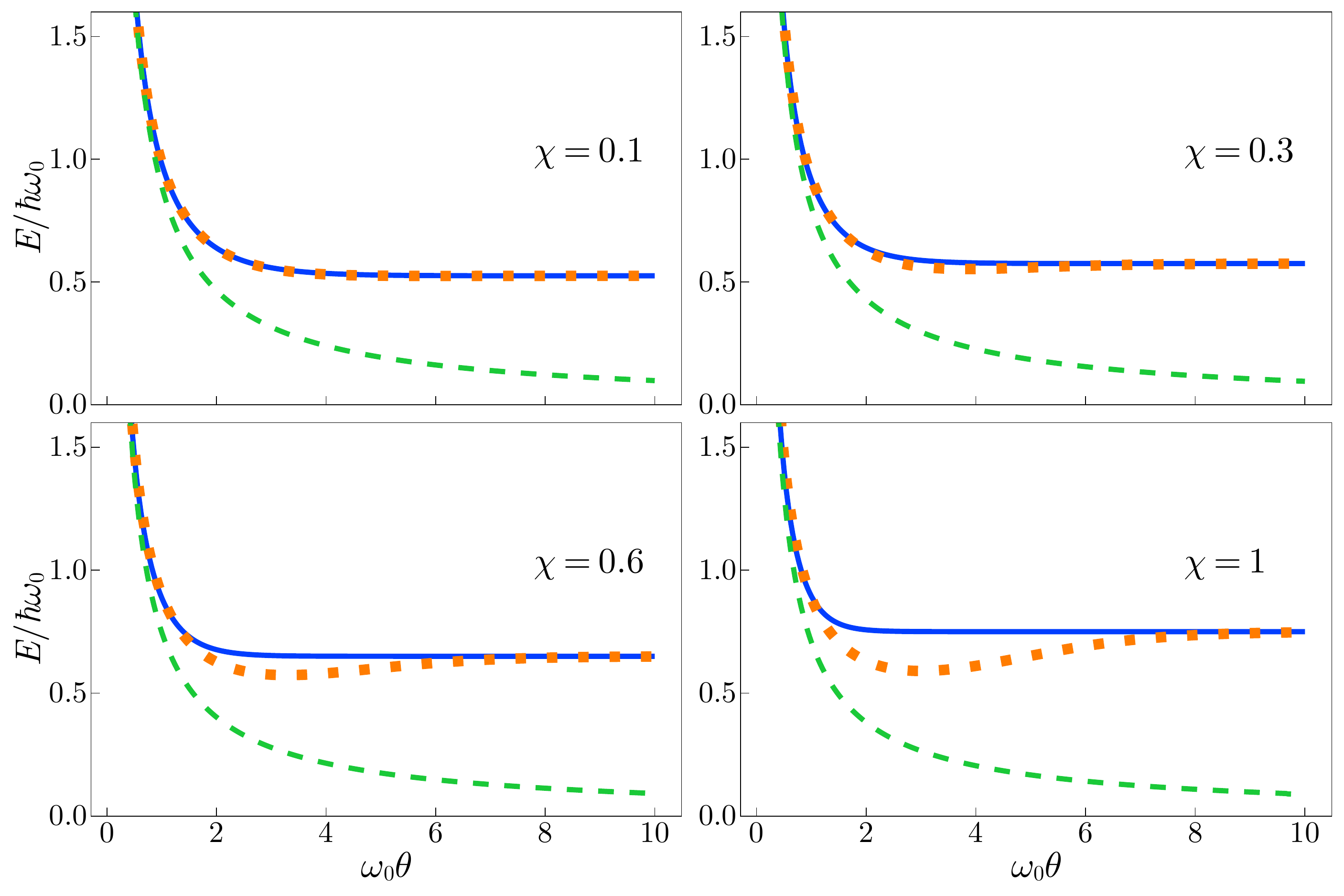}
    \caption{Average energy for the Kerr system as a function of thermal time. Solid blue line: quantum result; Orange markers: semiclassical approximation; Green dashed line: Classical result. }
    \label{energies kerr}
\end{figure}

In figure \ref{energies kerr}, we show the expectation value of the energy $E$ as a function of thermal time $\theta$ for different values of the parameter $\chi$.

\begin{figure}[ht]
    \centering
    \includegraphics[width=.95\linewidth]{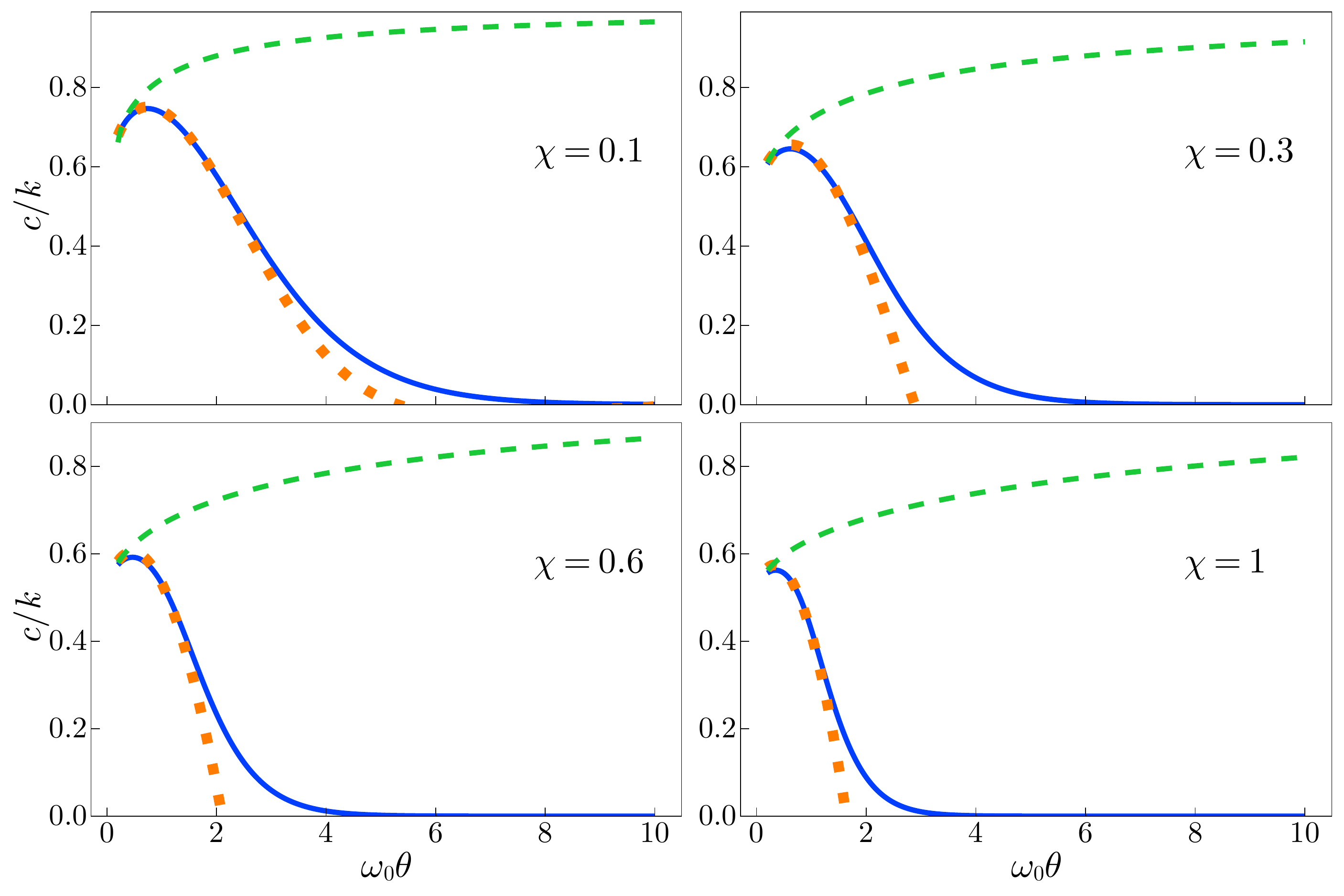}
    \caption{Specific heat for the Kerr system  as a function of thermal time. Solid blue line: quantum result; Orange markers: semiclassical approximation; Green dashed line: Classical result. }
    \label{heats kerr}
\end{figure}

In the canonical ensemble, the specific heat $c$ can be calculated in terms of the variance in the energy:
\begin{equation}
    c = k \beta^2 \left( \left< \hat{H}^2\right> - \left< \hat{H}\right>^2 \right).
\end{equation}
In this case, the Wigner symbol of the \textit{square} of the hamiltonian $H^2\left( \boldsymbol{x} \right)$ is significantly different from $\left [H\left( \boldsymbol{x} \right)\right]^2$. With this is mind, we show, in figure \ref{heats kerr}, the specific heat as a function of thermal time for different values of $\chi$.

As one sees, the quality of our semiclassical approximation is heavily dependent on the parameter $\chi$. As $\chi$ increases, we see a deviation from the quantum results for intermediate values of $\theta$, although, as foreseen, we have a good agreement at both high and low temperatures. 

\begin{figure}[ht]
    \centering
    \includegraphics[width=.8\linewidth]{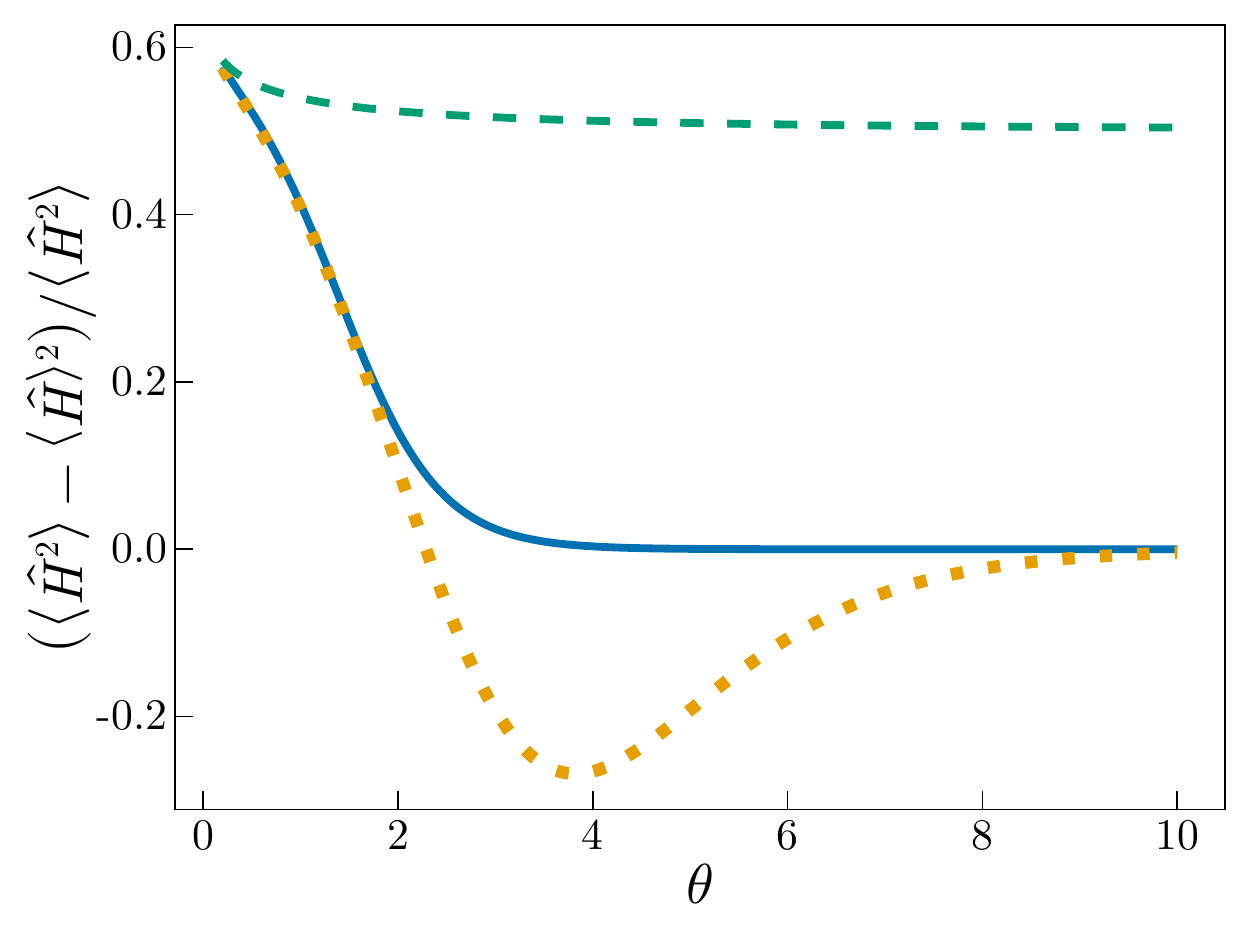}
    \caption{$\left(\left \langle \hat{H}^2\right \rangle  - \left \langle \hat{H}\right \rangle ^2\right) / \left \langle \hat{H}^2\right \rangle$ for $\chi = 0.5$. Solid blue line: quantum result; Orange markers: semiclassical approximation; Green dashed line: Classical result. }
    \label{heat analysis}
\end{figure}

It may seem somewhat disturbing that the semiclassical approximation does not respect the positivity of the heat capacity: the figure cuts off the negative
region. Yet it must be recalled that errors in $\left< \hat{H}^2\right>$ and
$\left< \hat{H}\right>^2$ may be added, whereas the averages are subtracted.
Indeed, Fig. \ref{heat analysis} shows that the variance in (57) is considerably smaller than $\left< \hat{H}^2\right>$. In general one may then expect the heat capacity to be a much more stringent test of approximations than the energy average, as will
occur further in our examples. One should note that a semiclassical version of the heat capacity as a second derivative of the partition function is not viable, as discussed in \cite{OZORIODEALMEIDA2021132951}. 

\section{Double Phase Space}
\label{Double Phase Space}

If we wish to apply our approximation for a broader class of systems, we must resort to numerical techniques. In this section, we restate the calculation of the terms in the integrand \eqref{trace of U O in terms of X} in a way that is well adapted for a numerical solution, and, in the following sections, we apply this method for a few more systems.

Instead of obtaining the centre $\boldsymbol x$ in equation \eqref{definition centre} propagating two trajectories, one forward and one backwards in time, it will be easier to calculate a single forward trajectory in a double phase space. In this new space, the centre $\boldsymbol x$ will play the role of position, while the conjugate momentum is given by $\boldsymbol{y} = \boldsymbol{J}\boldsymbol{\xi}$. The double hamiltonian \cite{de2009semiclassical,brodier2010complex,koda2015initial}
\begin{equation}
    \begin{aligned}
        \mathbb{H}(\boldsymbol{x},\boldsymbol{y}) &= H\left( \boldsymbol{x} - \frac{1}{2}\boldsymbol{J} \boldsymbol{y} \right) + H\left( \boldsymbol{x} + \frac{1}{2}\boldsymbol{J} \boldsymbol{y} \right) \\ &= H(\boldsymbol{x}_+) + H(\boldsymbol{x}_-) .
    \end{aligned}
\end{equation}
will then give the correct equations of motion, as one may check:
\begin{subequations}
\label{equações de movimento espaço de fase duplo}
    \begin{equation}
        \begin{aligned}
            \frac{\partial \mathbb{H}}{\partial \boldsymbol{x}} &= \nabla H(\boldsymbol{x}_+) + \nabla H(\boldsymbol{x}_-) \\ &= -\boldsymbol{J} \left( \dot{\boldsymbol{x}}_+-\dot{\boldsymbol{x}}_- \right) = -\boldsymbol{J} \dot{\boldsymbol{\xi}} = -\dot{\boldsymbol{y}};
        \end{aligned}
    \end{equation}
    \begin{equation}
        \begin{aligned}
            \frac{\partial \mathbb{H}}{\partial \boldsymbol{y}} &= \frac{\boldsymbol{J}}{2}\left[\nabla H(\boldsymbol{x}_+) - \nabla H(\boldsymbol{x}_-)\right] \\& = \frac{1}{2}\left( \dot{\boldsymbol{x}}_++\dot{\boldsymbol{x}}_- \right) =  \dot{\boldsymbol{x}}.
        \end{aligned}
    \end{equation}
\end{subequations}
For a given midpoint $\boldsymbol X$, these equations must then be solved under the initial conditions
\begin{equation}
\label{condições iniciais}
    \boldsymbol{x}\left( \boldsymbol{X},0 \right) = \boldsymbol{X}, \ \ \ \boldsymbol{y}\left( \boldsymbol{X},0 \right) = \boldsymbol{0}.
\end{equation}

In order to be able to define the trajectories for complex times, we simply promote the derivatives with respect to $t$ in \eqref{equações de movimento espaço de fase duplo} to derivatives with respect to a complex number $z$:
\begin{equation}
    \begin{aligned}
        \dfrac{d \boldsymbol{y}}{dz} &= -\dfrac{\partial \mathbb{H}}{\partial \boldsymbol{x}} \\
        \dfrac{d \boldsymbol{x}}{dz} &= \dfrac{\partial \mathbb{H}}{\partial \boldsymbol{y}}
    \end{aligned}
\end{equation}

If the functions $\boldsymbol x$ and $\boldsymbol y$ defined by these equations turn out to be analytic, then the line integral
\begin{equation}
    \label{first integral formula area}
    \Delta(z) = \int_\gamma \boldsymbol{y} \cdot \frac{d\boldsymbol{x}}{dz^\prime} \, dz^\prime ,
\end{equation}
which would be the extension of the area $\Delta$, given in \eqref{area}, for the complex plane, is only dependent on the endpoints of the path $\gamma$, which are $0$ and $z$.

\begin{figure}[ht]
    \includegraphics[width=.6\linewidth]{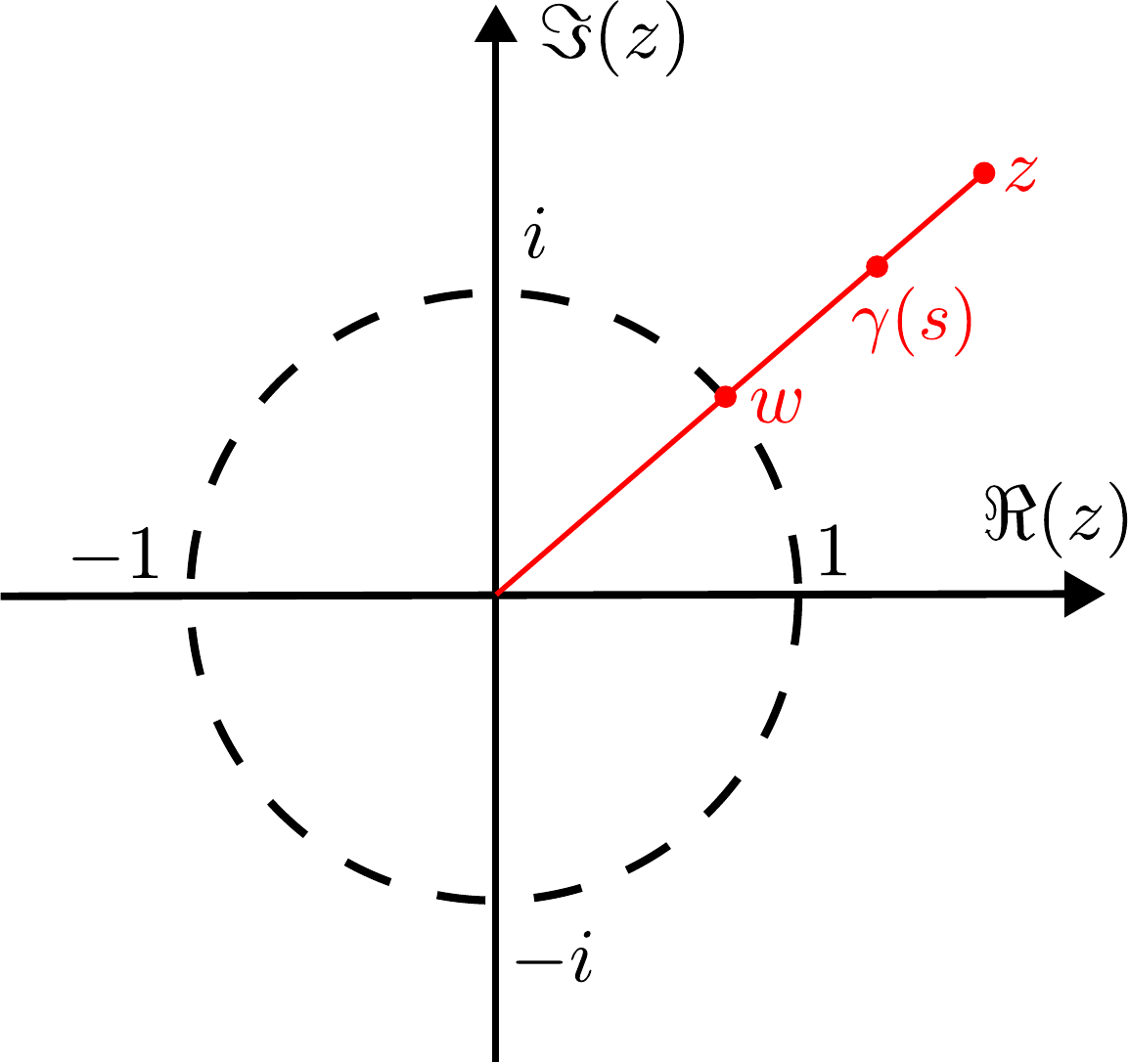}
    \centering
    \caption{Line segment joining the origin $0$ to $z$.}
    \label{complex plane}
\end{figure}

Assuming this is the case, we choose $\gamma$ to be the easiest path that joins $0$ and $z$ — the line segment — and parameterise it by the arc-length $s$. The explicit expression for the parametrization $\gamma(s)$ is then 
\begin{align*}
  \gamma \colon[0,|z|] &\to \mathbb{C}\\
  s &\mapsto  s w,
\end{align*}
where $w = z/|z|$. This path is illustrated in figure \ref{complex plane}.

Now, we compose $\boldsymbol{x}$ and $\boldsymbol{y}$ with $\gamma$, that is, we define the functions $\tilde{\boldsymbol{x}}:[0,|z|]\to \mathbb{C}, \ \tilde{\boldsymbol{x}}(s) = \boldsymbol{x}\circ \gamma(s) = \boldsymbol{x}(sw)$ and $\tilde{\boldsymbol{y}}:[0,|z|]\to \mathbb{C}, \ \tilde{\boldsymbol{y}}(s) = \boldsymbol{y}\circ \gamma(s) = \boldsymbol{y}(sw)$. By the chain rule, we deduce that these functions satisfy
\begin{equation}
    \begin{aligned}
        \dfrac{d \tilde{\boldsymbol{y}}}{ds} &= -w\dfrac{\partial \mathbb{H}}{\partial \boldsymbol{x}} \\
        \dfrac{d \tilde{\boldsymbol{x}}}{ds} &= w\dfrac{\partial \mathbb{H}}{\partial \boldsymbol{y}} 
    \end{aligned},
\end{equation}
which are \textit{almost} Hamilton's equations with real time $s$. If we define $\tilde{\boldsymbol{y}}_w = w^* \tilde{\boldsymbol{y}}$, where $^*$ denotes complex conjugation, and the modified hamiltonian
\begin{equation}
    \begin{aligned}
        \mathbb{H}_{w}\left(\boldsymbol{x},\boldsymbol{y}\right) &=  \mathbb{H}\left(\boldsymbol{x},w\boldsymbol{y}\right) \\ &= H\left( \boldsymbol{x} - \frac{w}{2}\boldsymbol{J} \boldsymbol{y} \right) + H\left( \boldsymbol{x} + \frac{w}{2}\boldsymbol{J} \boldsymbol{y} \right)
    \end{aligned}
\end{equation}
we, in fact, recover \textit{proper} Hamilton's equations
\begin{equation}
    \label{proper hamilton's equations}
    \begin{aligned}
        \dfrac{d \tilde{\boldsymbol{y}}_w}{ds} &= -\dfrac{\partial \mathbb{H}_{w}}{\partial \boldsymbol{x}} \\
        \dfrac{d \tilde{\boldsymbol{x}}}{ds} &= \dfrac{\partial\mathbb{H}_{w}}{\partial \boldsymbol{y}}
    \end{aligned}
\end{equation}
although with a hamiltonian that is generally complex, but that reduces to a real function if $w \in \left\{ 1,-1,i,-i \right\}$, which is the case for our interests ($w=-i$).

In terms of these quantities, the area is given by
\begin{equation}
\label{formula integral área}
    \begin{aligned}
        \Delta(z) &= w \int_{0}^{|z|/2} ds \ \tilde{ \boldsymbol{y}}_w(s) \cdot \frac{d\tilde{ \boldsymbol{x}}(s)}{ds}  \\
        &= w\int_{0}^{|z|/2} ds \ \tilde{ \boldsymbol{y}}_w(s) \cdot \left. \dfrac{\partial\mathbb{H}_{w}}{\partial \boldsymbol{y}} \right|_{\tilde{\boldsymbol{x}}(s),\tilde{\boldsymbol{y}}_w(s)}
    \end{aligned},
\end{equation}
which, alternatively, may be cast in the form of a initial value problem
\begin{equation}
    \label{general area}
    \begin{aligned}
        \frac{d\Delta}{ds} &= w \tilde{ \boldsymbol{y}}_w(s) \cdot \left. \dfrac{\partial\mathbb{H}_{w}}{\partial \boldsymbol{y}} \right|_{\tilde{\boldsymbol{x}}(s),\tilde{\boldsymbol{y}}_w(s)} \\
        \Delta(0) &= 0
    \end{aligned}
\end{equation}

The last missing ingredient is a way to calculate the jacobian $\partial \boldsymbol x/ \partial \boldsymbol X$. It can be obtained by differentiating \eqref{proper hamilton's equations} with respect to $\boldsymbol X$, which give us the equations
\begin{subequations}
\label{jacobianos}
    \begin{equation}
        \frac{d}{ds} \frac{\partial\tilde{ \boldsymbol{y}}_w}{\partial \boldsymbol{X}} = - \frac{\partial^2\mathbb{H}_w}{\partial \boldsymbol{x} \partial \boldsymbol{y}}\frac{\partial\tilde{ \boldsymbol{y}}_w}{\partial \boldsymbol{X}} - \frac{\partial^2\mathbb{H}_w}{\partial \boldsymbol{x}^2}\frac{\partial \tilde{\boldsymbol{x}}}{\partial \boldsymbol{X}};
    \end{equation}
    \begin{equation}
        \frac{d}{ds} \frac{\partial \tilde{\boldsymbol{x}}}{\partial \boldsymbol{X}} = \frac{\partial^2\mathbb{H}_w}{\partial \boldsymbol{y}^2}\frac{\partial\tilde{ \boldsymbol{y}}_w}{\partial \boldsymbol{X}} + \frac{\partial^2\mathbb{H}_w}{\partial \boldsymbol{x} \partial \boldsymbol{y}}\frac{\partial \tilde{\boldsymbol{x}}}{\partial \boldsymbol{X}} ,
    \end{equation}
\end{subequations}
solved under the initial conditions
\begin{equation}
    \frac{\partial\tilde{ \boldsymbol{y}}_w}{\partial \boldsymbol{X}}\biggr|_{\theta = 0} = \boldsymbol{0}; \ \ \ \frac{\partial \tilde{\boldsymbol{x}}}{\partial \boldsymbol{X}}\biggr|_{\theta = 0} = \boldsymbol{I}.
\end{equation}

In order to calculate \eqref{trace of U O in terms of X} with $t = -i\theta$, one must then solve the initial value problems specified by \eqref{proper hamilton's equations}, \eqref{general area} and \eqref{jacobianos} with $w=-i$, and obtain the solution at $s = \theta/2$. If we have $d$ degrees of freedom, this will be a system of coupled differential equations with $8d^2+4d + 1$ \textit{real} variables. For the remaining of this work, we will test the procedure here described for the cases $d=1,2$.

It should be noted that the definition of a real double Hamiltonian in replacement of a complex Hamiltonian is not unique. Our construction, specially suited to the Wigner-Weyl representation, coincides neither with the double Hamiltonians designed for propagating coherent states by de Aguiar et al. \cite{de2005semiclassical}, nor with the double Hamiltonian for the Boltzmann operator of Yan and Shao \cite{yan2019semiclassical}. Indeed, the double trajectories in this reference decouple into pairs of simple trajectories, due to the assumption of a quadratic momentum dependence. 

For each choice of decomplexification one obtains different trajectories in their own phase space. Unlike the simple equivalence of the various variants related directly by Fourier transforms, the SC equivalence for decomplexified Hamiltonians is, so far, a question which relies on numerical investigaton. A mixed position-momentum representation is employed in \cite{yan2019semiclassical} to derive their semiclassical approximation of the density matrix, but it requires a further Fourier transform (beyond the Wigner transform) to relate their phase space formulae to the thermal Wigner function. It is interesting that a composition of Herman-Kluck propagators evolving for positive and negative half-times is also proposed, but there it is only optional, in contrast to its essential role within the present more general theory,

\section{Morse System}
\label{Morse System}

The Morse potential \cite{PhysRev.34.57} is given by the expression
\begin{equation}
    V(r) = D \left[ 1-e^{-a(r-r_e)} \right]^2,
\end{equation}
where $r$ is a radial coordinate, $D$ is the dissociation energy, $r_e$ is the equilibrium distance and $a$ is a constant with dimensions of inverse distance. 

\begin{figure}[ht]
    \centering
    \includegraphics[width=.6\linewidth]{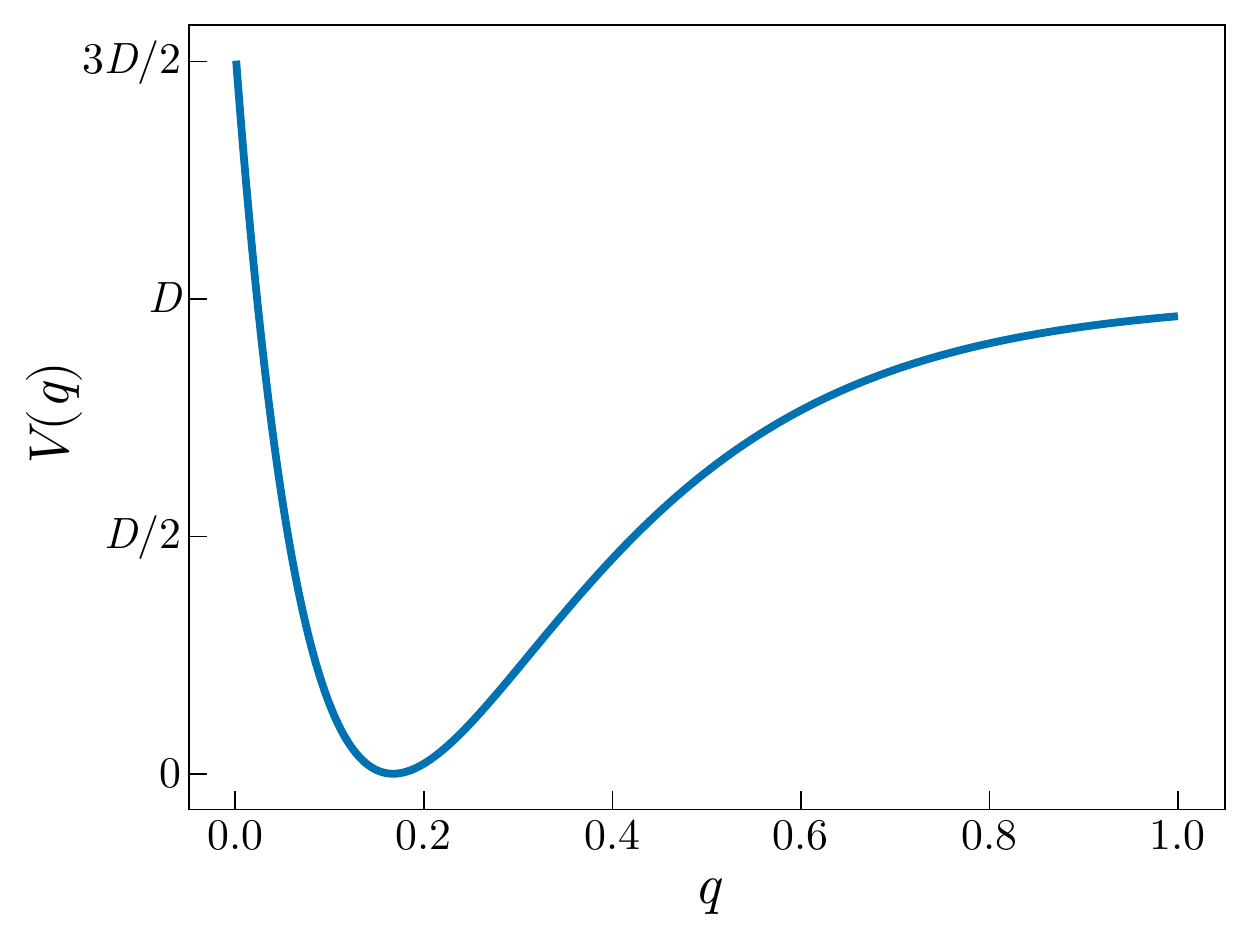}
    \caption{Morse potential.}
    \label{morse potential fig}
\end{figure}

It is a model for the vibration of diatomic molecules, that takes into account the possibility of the dissociation of the bond.

As for every one dimensional system without an explicit dependence on time, the trajectories of a particle under the action of the Morse potential can be obtained by quadrature. We will explore a few insights that can be given by these expressions, but they are too cumbersome to actually use in the calculation of thermodynamic averages. For that, we will resort to the method described in the previous section.

In order to more easily describe these trajectories, it will be convenient to introduce the dimensionless coordinate $q = a(r-r_e)$ and the frequency
\begin{equation}
    \omega = \sqrt{\frac{2Da^2}{m}}.
\end{equation}
In this way, the lagrangian for this system is
\begin{equation}
    L = \frac{m \dot{r}^2}{2} - V(r) = D \left[ \left( \frac{\dot{q}}{\omega} \right)^2 - \left( 1-e^{-q} \right)^2 \right],
\end{equation}
from which we obtain a conjugate momentum
\begin{equation}
    p = \frac{\partial L}{\partial \dot{q}} = \frac{2D\dot{q}}{\omega^2}
\end{equation}
and a hamiltonian
\begin{equation}
    H = p \dot{q} - L = \frac{\omega^2p^2}{4D} + D\left(1-e^{-q}\right)^2.
\end{equation}

The solutions of Hamilton's equations are then \cite{slater1957classical,PhysRevA.37.796}
\begin{equation}
    \begin{aligned}
        q(t) &= \ln \left[ \frac{1-\sqrt{\epsilon} \cos\left(\Omega t + \phi \right)}{1-\epsilon} \right]; \\
        p(t) &= \frac{2D \sqrt{\epsilon(1-\epsilon)}}{\omega}\frac{\sin\left(\Omega t + \phi \right)}{1-\sqrt{\epsilon} \cos\left(\Omega t + \phi \right)} .
    \end{aligned} 
\end{equation}
Here
\begin{equation}
    \epsilon = \frac{E}{D} = \left( \frac{\omega p}{2D}\right)^2 + \left(1-e^{-q}\right)^2 < 1
\end{equation}
is the orbit's normalized energy, $\Omega = \sqrt{1-\epsilon} \ \omega$ is the orbit's frequency and $\phi$ is a phase determined by the initial conditions.

We see that, for $t \in \mathbb{R}$, we have $\sqrt{\epsilon} \cos\left(\Omega t + \phi \right) < 1$, and the orbits are well behaved. Nonetheless, if we allow $t \in \mathbb{C}$, we do not have this guarantee. Indeed, by taking $t = -i s, \ s \in \mathbb{R}$, and restricting ourselves to initial conditions of the form $p_0=0$ and $q_0<0$, which imply that $\phi = 0$, we obtain
\begin{equation}
    \begin{aligned}
        q(t) &= \ln \left[ \frac{1-\sqrt{\epsilon} \cosh\left(\Omega s \right)}{1-\epsilon} \right];  \\ p(t)& = \frac{2D \sqrt{\epsilon(1-\epsilon)}}{\omega}\frac{i\sinh\left(\Omega s \right)}{1-\sqrt{\epsilon} \cosh\left(\Omega s \right)},
    \end{aligned}
\end{equation}
and we see that the trajectories diverge at a finite critical time $s_c$ that satisfies $1-\sqrt{\epsilon} \cosh\left(\Omega s_c \right)=0$, or, choosing the positive solution, one arrives, explicitly, at
\begin{equation}
    \omega s_c = \frac{1}{\sqrt{1-\epsilon}} \ln \left( \frac{1}{\sqrt{\epsilon}} + \sqrt{\frac{1}{\epsilon}-1} \right).
\end{equation}

\begin{figure}[ht]
    \centering
    \includegraphics[width=.6\linewidth]{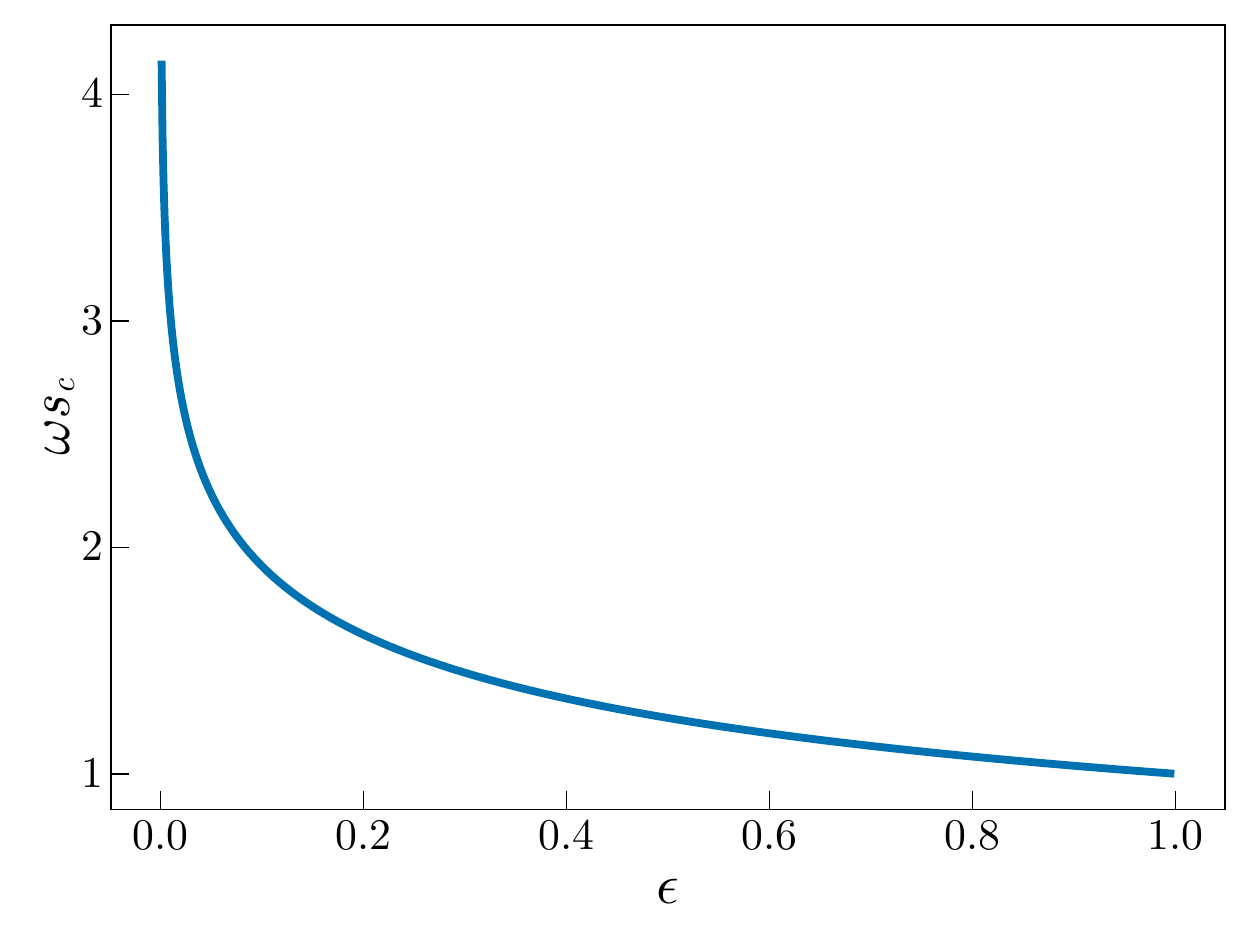}
    \caption{Critical time as a function of energy.}
    \label{critical time}
\end{figure}

According to figure \ref{critical time}, one sees that $s_c(\epsilon)$ is a decreasing function that satisfies
\begin{equation}
    \lim_{\epsilon \to 0} \omega s_c = \infty; \ \ \ \lim_{\epsilon \to 1} \omega s_c = 1.
\end{equation}
In this way, we obtain divergent trajectories when $\omega s > 1$, which start at the region with $\epsilon\to 1$ or, equivalently, with $q_0 \to -\ln 2$, and advance towards $\epsilon = q_0 =0$.

These divergent trajectories are, in a first moment, a disaster for our theory. Note that the integrand of \eqref{first integral formula area} has a pole in this case, and therefore the integral is path dependent, which translates to multiple branches for the area $\Delta$. It is then not clear which branch to choose. Beyond that, one may see, numerically, that these divergent trajectories are accompanied by the appearance of caustics. Furthermore, if one directly applies the method described in the previous section, it is possible to obtain good results until a thermal time $\theta = 2$ (remember that the trajectories are evolved until $s$ reaches $\theta/2$), when, suddenly, the approximation fails enormously. The culprit appears to be the fact that, after the caustic, the euclidean action rapidly grows from very negative to very positive values, which translates to a large growth in the integrand \eqref{trace of U O in terms of X}. The fact is that, while there is a good understanding of caustic traversals for real time \cite{de2014metaplectic}, the same cannot be said for imaginary times. Note that even the change of coordinates $\boldsymbol x \mapsto \boldsymbol X$ may fail, as the jacobian stops being invertible. 

Nonetheless, the simple trick of \textit{discarding} the trajectories that cross caustics, imposing that they should not contribute to the integral \eqref{trace of U O in terms of X}, seems to completely eliminate our problem. This question certainly deserves further investigation, but, for now, we will stick to this \textit{ad hoc} trick, as it appears to work very well in practice.

The comparison with the quantum result is also straightforward for the Morse system, as its quantum version is well understood — there are a finite number of bound eigenstates, whose eigenfunctions and eigenenergies are known \cite{dahl1988morse}. By introducing the dimensionless parameter
\begin{equation}
    \chi = \frac{\hbar \omega}{4D},
\end{equation}
we may write the eigenenergies corresponding to the bound states as
\begin{equation}
    \label{energy spectrum morse}
    E_n = \hbar \omega \left[ \left( n + \frac{1}{2} \right) - \chi \left( n + \frac{1}{2} \right)^2 \right], \ \ \ n = 0,1, \ldots, N
\end{equation}
where
\begin{equation}
\label{numero de estados ligados}
    N=\left \lfloor{\frac{1}{2\chi} - \frac{1}{2}}\right \rfloor, 
\end{equation}
and denotes $\left \lfloor{x}\right \rfloor $ the largest integer greater than or equal to $x$.

In order for us to have an idea of the order of magnitude in physically relevant cases, we exhibit, on table \ref{dados moleculares}, the values of $\chi$ and $N$ that best fit experimental results for molecules of hydrogen, oxygen and nitrogen.

\begin{table}[ht]
\caption{Values of $\chi$ and $N$ for some molecules. Calculated from \cite{haynes2016crc}.}\label{dados moleculares}%
\begin{tabular}{@{}ccc@{}}
\toprule
Molecule  & $\chi$ & $N$\\
\midrule
$H_2$   & $2.76 \times 10^{-2}$ & $17$\\

 $O_2$&   $7.58 \times 10^{-3}$  & $65$  \\

 $N_2$ & $6.07 \times 10^{-3}$  & $81$  \\
\botrule
\end{tabular}

\end{table}

This system has the peculiarity of presenting both bound and free states. In this scenario, one often is only preoccupied with the regime in which the temperature is bellow the value of excitation for the free states and, therefore, a regime in which they will not contribute to the thermodynamics of the system. In this case, the classical prescription is to only perform the integrations defining the thermodynamic quantities in the region of phase space corresponding to bound states \cite{riganelli2001rovibrational}. We will stick with this classical prescription in the semiclassical case, therefore restricting the integration region in \eqref{trace of U O in terms of X}.

\begin{figure}[ht]
    \centering
    \includegraphics[width=.95\linewidth]{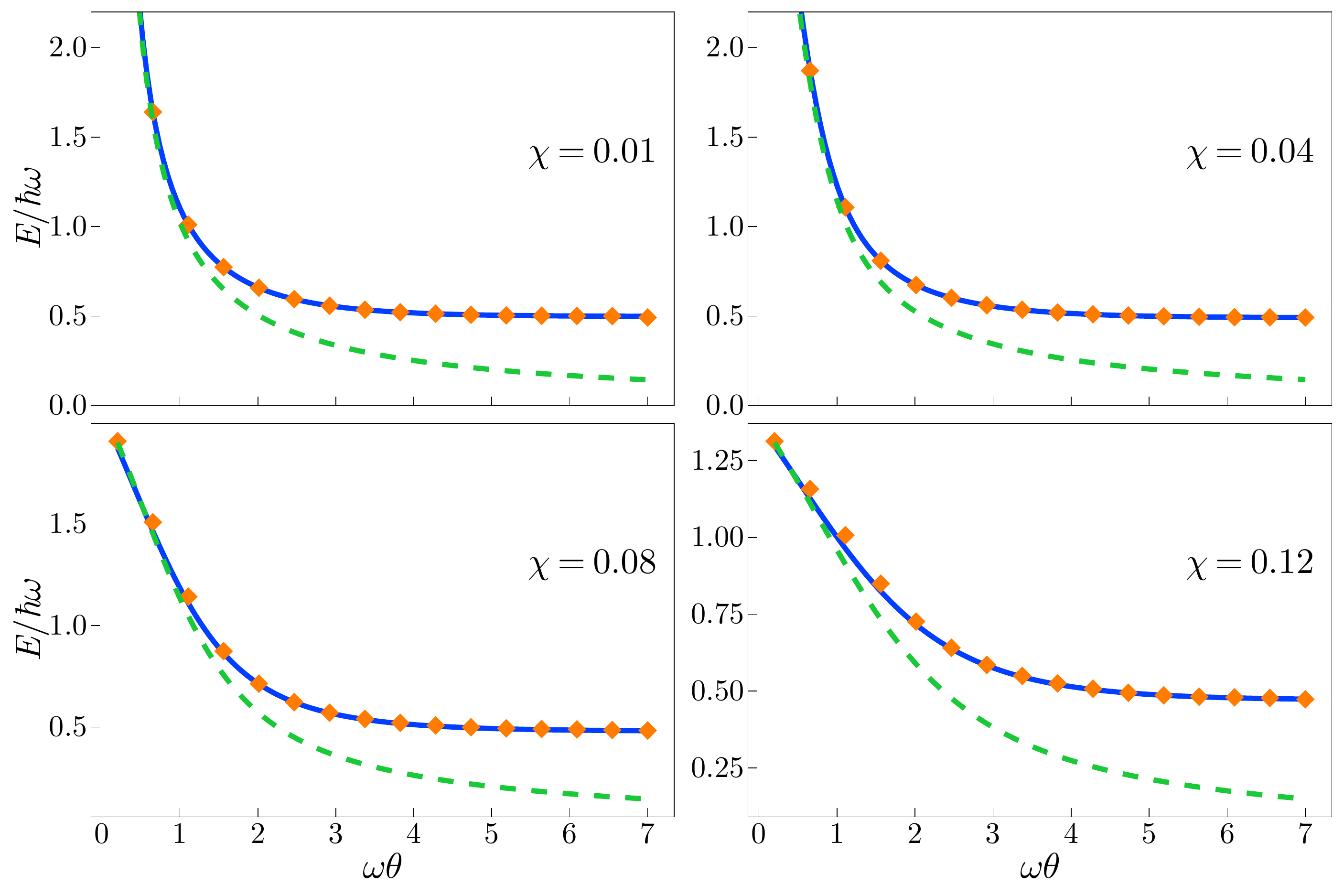}
    \caption{Average energy for the Morse system  as a function of thermal time. Solid blue line: quantum result; Orange markers: semiclassical approximation; Green dashed line: Classical result. }
    \label{energies morse}
\end{figure}

\begin{figure}[ht]
    \centering
    \includegraphics[width=.95\linewidth]{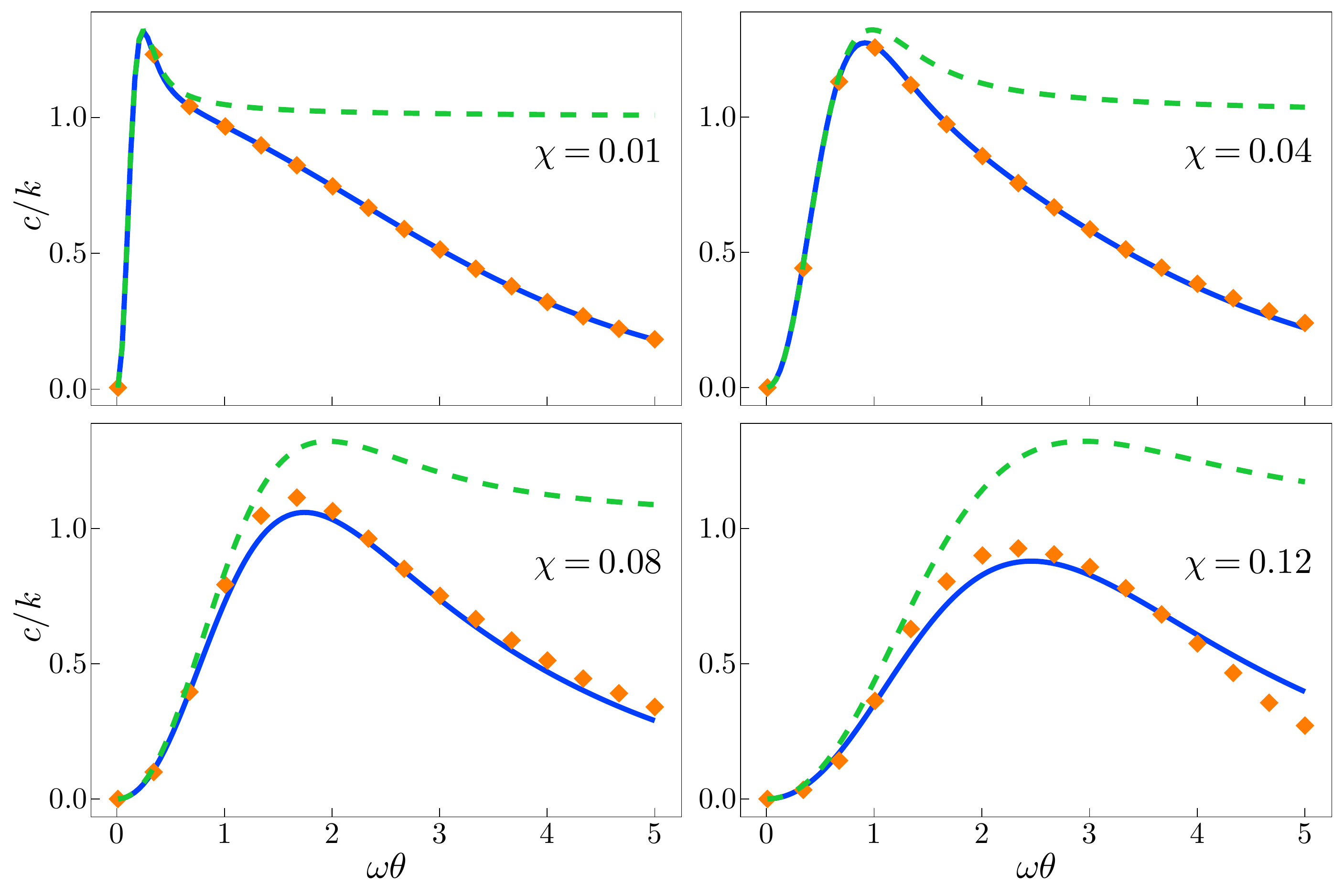}
    \caption{Specific heat for the Morse system as a function of thermal time. Solid blue line: quantum result; Orange markers: semiclassical approximation; Green dashed line: Classical result. }
    \label{heats morse}
\end{figure}

We then repeat the analysis done for the Kerr system, by calculating the energy and the heat capacity given by our approximations, and comparing them with the classical and the quantum cases. The results can be seen in figures \ref{energies morse} and \ref{heats morse}.

The values given by our approximation are quite remarkable, specially in the case of the energy, where one sees almost no deviation from the quantum result. We note that, for $\chi = 0.12$, we have only four bound states. The values of the heat capacity are less accurate, as also happened in the Kerr system, but are still far superior to the classical case. We also note that the approximation seems to fare better the lower the value of $\chi$. This may be related to the fact that, when $\chi \to 0$, we recover the spectrum of the harmonic oscillator, as can be seen from \eqref{energy spectrum morse}.

\begin{figure}[ht]
    \centering
    \includegraphics[width=.95\linewidth]{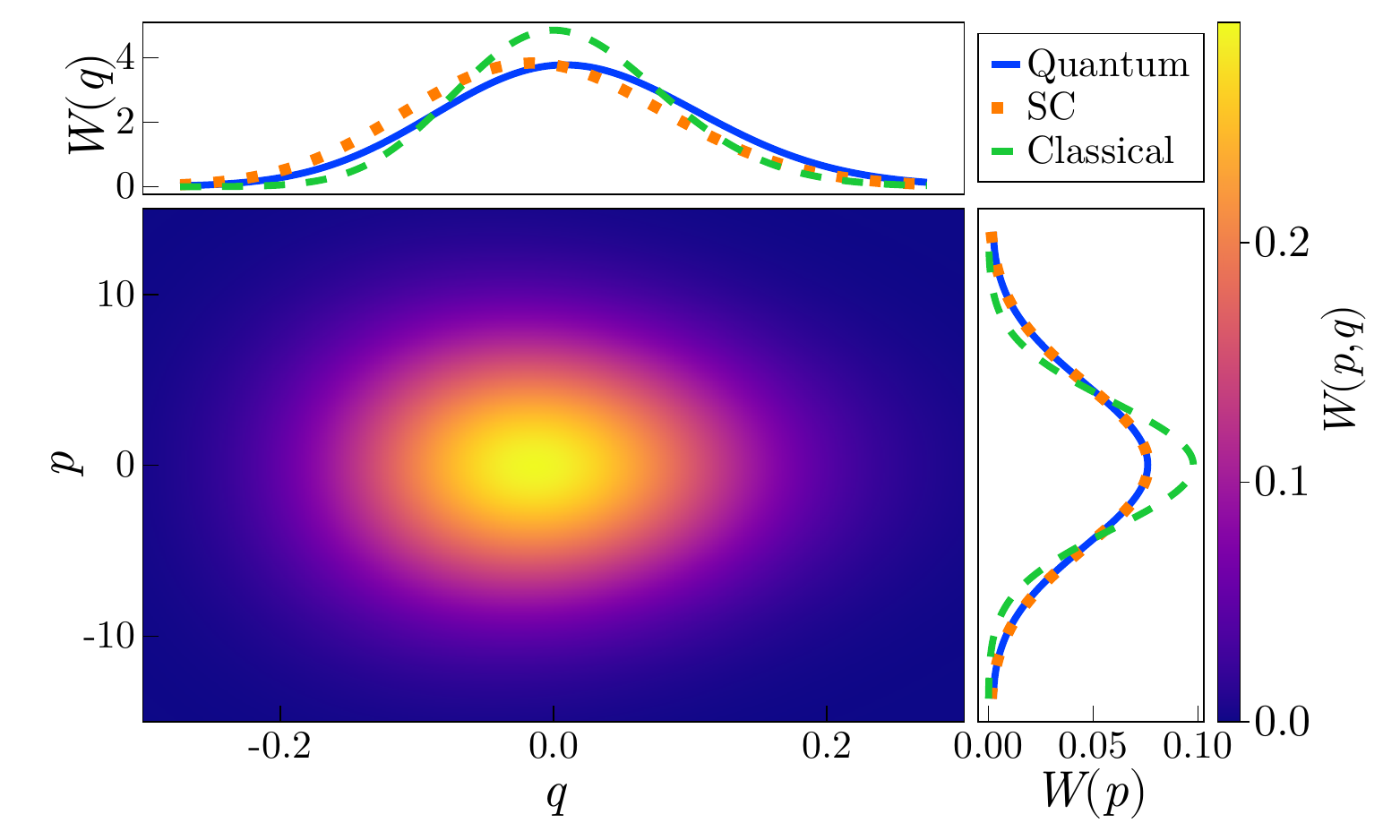}
    \caption{The Semiclassical Wigner function is shown in the heat map. Different versions of the projections $W(p)$ and $W(q)$ are shown: Solid blue line: quantum result; Orange markers: semiclassical approximation; Green dashed line: Classical result, obtained using the classical Boltzmann distribution. }
    \label{wigner morse}
\end{figure}

Because of the change of variables $\boldsymbol x \mapsto \boldsymbol X$, which results in formula \eqref{trace of U O in terms of X}, we only have access to a displacement of the thermal Wigner function. Nonetheless, remembering that, according to \eqref{definition wigner}, the Wigner function $W(\boldsymbol x^\prime)$ is proportional to $\left< \hat{R}_{\boldsymbol{x}^\prime} \right>$, and using the fact that $\hat{R}_{\boldsymbol{x}^\prime}(\boldsymbol{x}) = \delta\left( \boldsymbol{x} - \boldsymbol{x}^\prime\right)$, we may write
\begin{equation}
    \begin{aligned}
        W(\boldsymbol x^\prime) &\propto \int d \boldsymbol{X} \left| \frac{\partial \boldsymbol x}{\partial \boldsymbol X}\right|^{1/2} \exp\left[\Delta(\boldsymbol{X}) /\hbar - \beta H(\boldsymbol{X})\right]  \delta\left( \boldsymbol{x} - \boldsymbol{x}^\prime\right) \\ &= \left . \left| \frac{\partial \boldsymbol x}{\partial \boldsymbol X}\right|^{1/2} \exp\left[\Delta(\boldsymbol{X}) / \hbar - \beta H(\boldsymbol{X})\right] \right|_{\boldsymbol{X} = \boldsymbol{X}^\prime}
    \end{aligned}
\end{equation}
where $\boldsymbol{X}^\prime$ is the midpoint which gets mapped to $\boldsymbol{x}^\prime$ under equations \eqref{proper hamilton's equations}. One may also characterize $\boldsymbol{X}^\prime$ as the zero of the function
\begin{equation}
    f(\boldsymbol{X}) = \boldsymbol{x}\left( \boldsymbol{X} \right) - \boldsymbol{x}^\prime. 
\end{equation}
Assuming that no caustics have been traversed, this zero is unique, and may be found by a standard Newton-Raphson method, allowing us to calculate $W(\boldsymbol x^\prime)_{SC}$ as well as the marginal distributions
\begin{subequations}
    \begin{equation}
        W(p) = \int dq W(p,q)
    \end{equation}
    \begin{equation}
        W(q) = \int dp W(p,q)
    \end{equation}
\end{subequations}
which are the expectation values of the operators $\left|p\right>\left<p\right|$ and $\left|q\right>\left<q\right|$, respectively. These results for the Morse system, with a thermal time $\omega \theta = 3$ and $\chi = 0.01$ are shown in figure \ref{wigner morse}. We also show the quantum projections, as well as the classical one, which is obtained from the classical Boltzmann distribution.

It is possible to see that our semiclassical approximation for $W(p)$ is essentially exact, while $W(p)$ appears to be a displaced version of its quantum counterpart.

\section{Nelson System}
\label{Nelson System}

\begin{figure}[ht]
    \centering
    \includegraphics[width=.5\linewidth]{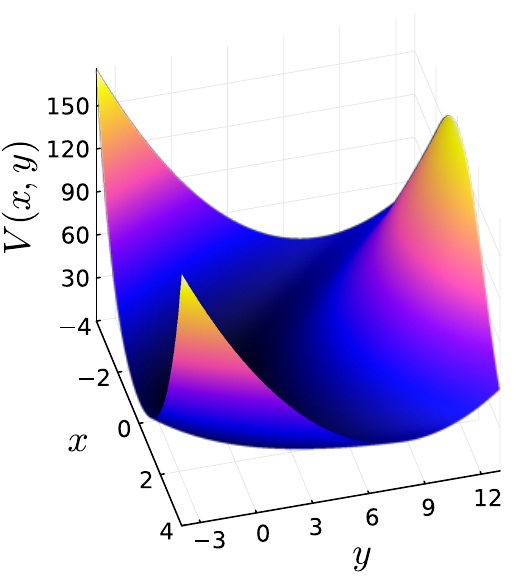}
    \caption{Nelson potential with $\mu = 2$}
    \label{nelson potential fig}
\end{figure}

The Nelson system \cite{baranger1987periodic} is described by a hamiltonian
\begin{equation}
    H(\mathbf{x}) = \frac{1}{2}\left(p_x^2 + p_y^2 \right) + V(x,y)
\end{equation}
which represents a particle of unit mass in two dimensions under the action of the Nelson potential
\begin{equation}
    V(x,y) = (x^2/2-y)^2 + \mu x^2
\end{equation}
where $\mu$ is a parameter. The potential is illustrated in figure \ref{nelson potential fig}.

\begin{figure}[ht]
    \centering
    \includegraphics[width=\linewidth]{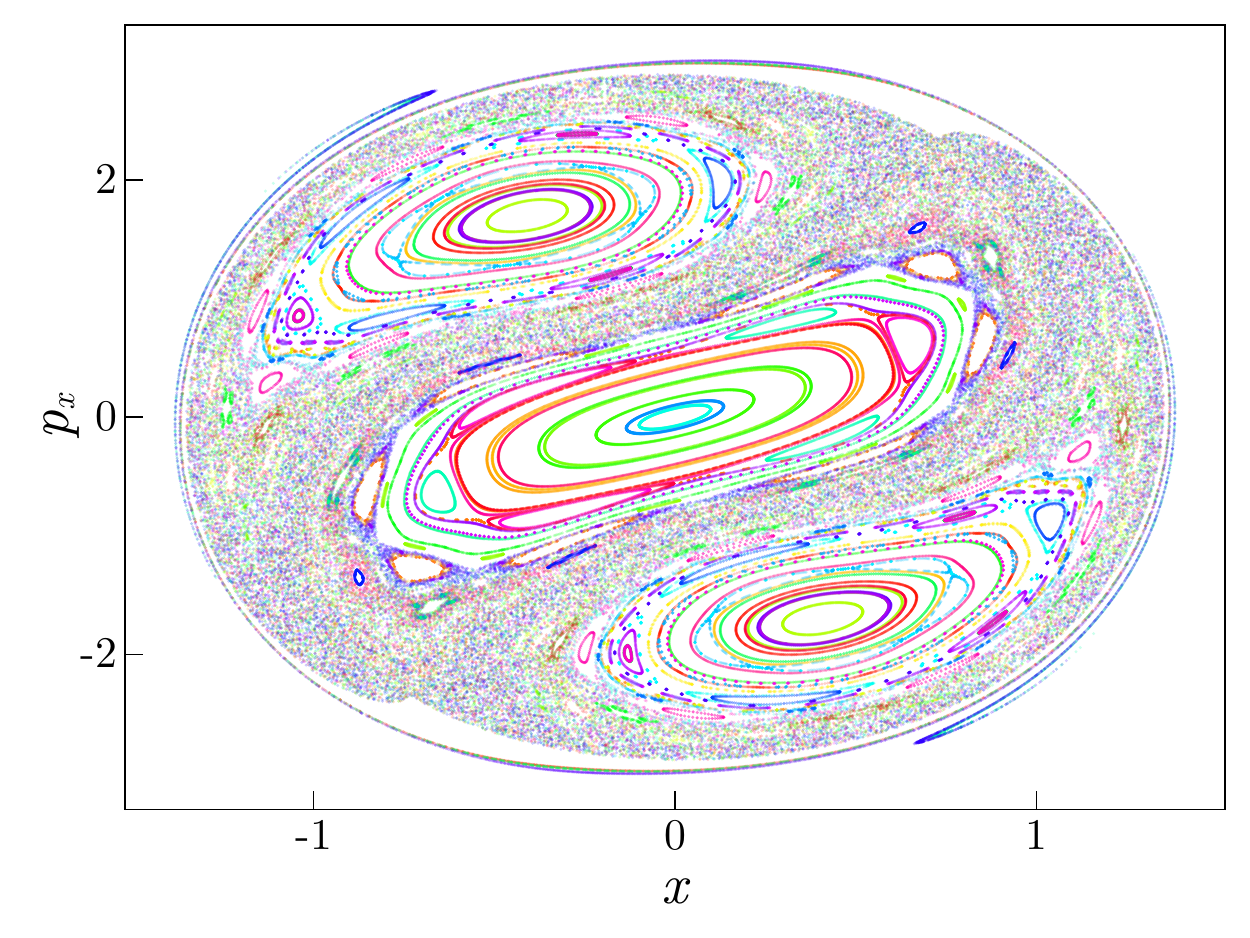}
    \caption{Poincaré section for the Nelson system with $\mu = 2$ and energy $E = 4.8$ with respect to the hyperplane $y=0$. Each color represents a different trajectory.}
    \label{poincare nelson}
\end{figure}

The classical dynamics in real time exhibits a generic mixture of stable regions within a chaotic sea as exemplified by its Poincaré section, visualized in Figure \ref{poincare nelson}. Bifurcation trees of its periodic orbits have
been intensively studied in  \cite{provost1995signatures,provost1995uniform,ribeiro2004semiclassical}. The important point to be borne in mind is that the range
between regular (integrable) and fully chaotic classical motion pertains to the infinite time limit. For finite time, the solutions of the Hamilton-Jacobi equation,
which are the backbone of semiclassical approximations of finite time quantum evolution, make no qualitative distinction between these alternatives. In any case, it is reassuring that our full double hamiltonian formalism is successful even for the thorniest types of generic mixed systems.  

Here, we simply use the Nelson system as an example of an application of our methods for a nontrivial system in two degrees of freedom. We then consider an ensemble of particles of unit mass under the action of this potential and calculate the thermodynamic averages associated with this system.

\begin{figure}[ht]
    \centering
    \includegraphics[width=.95\linewidth]{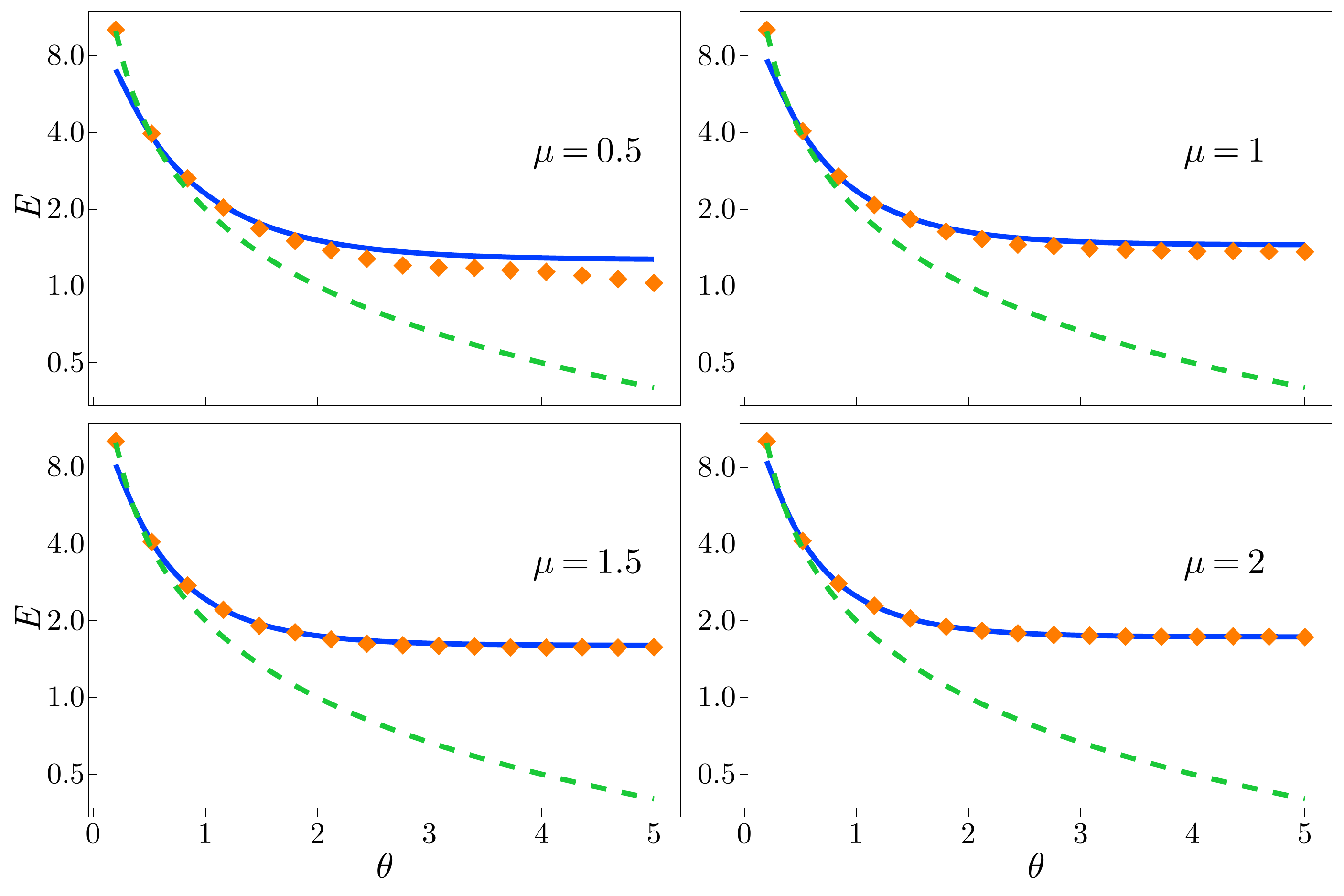}
    \caption{Average energy for the Nelson system as a function of thermal time. Solid blue line: quantum result; Orange markers: semiclassical approximation; Green dashed line: Classical result. }
    \label{nelson energies fig}
\end{figure}

\begin{figure}[ht]
    \centering
    \includegraphics[width=.95\linewidth]{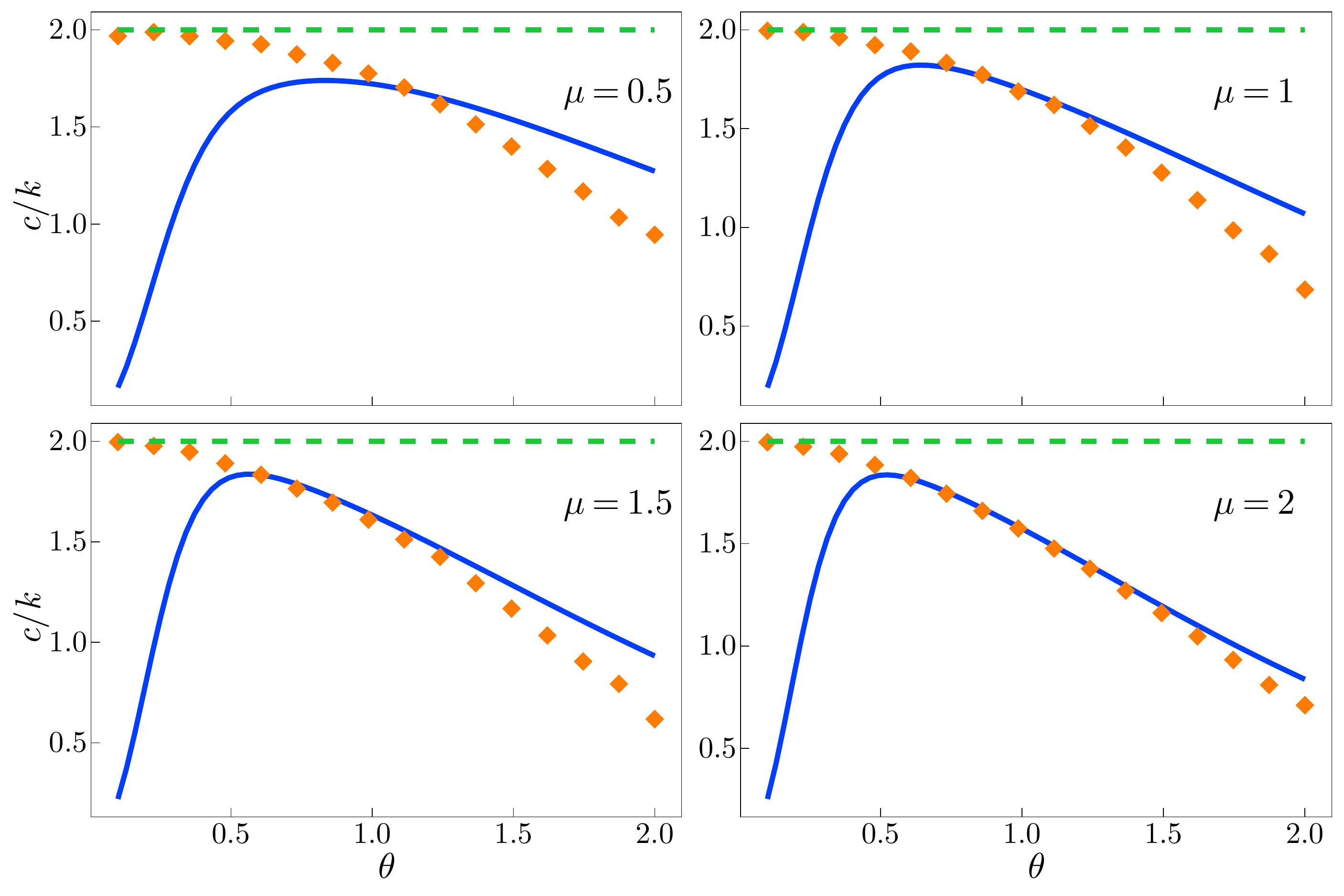}
    \caption{Specific heat for the Nelson system as a function of thermal time. Solid blue line: quantum result; Orange markers: semiclassical approximation; Green dashed line: Classical result. }
    \label{heat nelson}
\end{figure}

In this case, there are no analytical formulas available, and we must resort to numerical techniques in order to calculate the quantum energy spectrum. Because of this limitation, we only have access to a finite number of levels, and only the lower ones are reliable. This, in turn, will only give a reliable approximation for the thermodynamic quantities in the low temperature limit. On the other hand, the classical framework, as it is known, will give good results in the high temperature limit. Our hope is that our semiclassical approximation can join these two extremes in a satisfactory manner.

The semiclassical approximation for the energy, shown in figure \ref{nelson energies fig}, seems to bridge very well the high temperature limit, given by the classical result, and the low temperature one, given  by the quantum result. Unfortunately, the approximation for the heat capacity seems to fail for much smaller values of $\theta$, specially for $\mu = 0.5$, as can be seen in figure \ref{heat nelson}.

\section{Discussion}

Having reviewed and incremented the semiclassical approximation for the thermal Wigner function, we developed it into a numerical method that can be applied to a broad class of systems, including Hamiltonians that are not quadratic in their momenta. Even though further investigation will be required, we obtained good agreement of energy averages with the quantum results for a wide range of different systems and parameters. It is presumed that that this method can be useful for systems with many degrees of freedom, as its quadratic scaling law can keep  computation tractable even for high dimensions.

The Weyl propagator employed here belongs to the class of propagators related to the original Van Vleck propagator by various Fourier transforms:  All of these require the so called 'root search'. Thus, whereas one seeks trajectories with given end positions for the Van Vleck propagator, it is the centre point between the extremities of the trajectory that is prescribed in the Wigner-Weyl representation. So it is only for integrals involving the semiclassical Weyl propagator (or its Wick rotation) that one can switch from the centre to the initial or final value of the trajectory (as in \cite{de2013initial}) or to its midpoint, in the present instance. 

Each representation illuminates a different aspect of quantum mechanics. Thus, the position representation, so far favoured by the vast majority of computations in this field \cite{yan2019semiclassical}, can easily provide the probability density for positions, that is, the diagonal matrix elements of the density matrix are supplied explicitly. On the other hand, the momentum density requires a double Fourier transform over the pair of positions. In contrast, the Wigner function provides either density by a simple projection integral \cite{de1998weyl}, as exemplified by the momentum and position densities for the Morse system, shown in Fig. \ref{wigner morse}. 

A further unique feature of the Wigner-Weyl representation is its symplectic invariance \cite{de2014metaplectic}, that is, unitary quantum (metaplectic) transformations corresponding to classical linear canonical phase space transformations transport the Wigner function classically. Thus, the Wigner function supplies through simple projection integrals, not only the momentum probability density and the position probability density, but the probability density along any Lagrangian plane in phase space, that is, any plane where the action for any closed circuit is null. This multiple probability content of the density operator can be used to reconstruct it through multiple measurements in the process of quantum tomography \cite{PhysRevA.40.2847}, but prior knowledge of the thermal Wigner function is welcome shortcut.

\section*{Acknowledgments}
	
We thank Gabriel Lando for his advice on the numerics. Funding provided by Conselho Nacional de Desenvolvimento Científico e Tecnológico (CNPq) and Instituto Nacional de Ciência e Tecnologia de Informação Quântica is gratefully acknowledged.

\section*{Declarations}
All authors certify that they have no affiliations with or involvement in any organization or entity with any financial interest or non-financial interest in the subject matter or materials discussed in this manuscript

\section*{Data Availability}
The code used in this article can be found at \cite{SCCanonicalEnsemble}.

\begin{appendices}
    \section{Wigner symbol of normal forms}
\label{Símbolo de Wigner para formas normais}

In order to calculate the Wigner symbol of an operator of the form \eqref{quantum normal form}, it is sufficient to do so for the monomials $\hat{o}^n$, where $\hat{o} = \hat{p}^2+\hat{q}^2$. Our strategy will consist in finding a recurrence relation that allows us to calculate $o^{n+1}(p,q)$ in terms of $o^n(p,q)$. As the initial term $o^1(p,q) = o(p,q) = p^2+q^2$ is readily obtained, the problem is solved.

For that, we first observe that, as $\hat{o}^n$ is hermitian, $o^n(p,q)$ must be real. Using this fact, writing $\hat{o}^{n+1} = \hat{o} \hat{o}^{n}$, and applying Groenewold's rule \eqref{groenewold}, we arrive at the recurrence relation
\begin{equation}
\label{recorrencia1}
    o^{n+1}(p,q) = \left( p^2+q^2 - \frac{\hbar^2}{4}\nabla^2 \right)o^n(p,q)
\end{equation}
where $\nabla^2 = \partial_p^2 + \partial_q^2$. This relation is further simplified if we introduce the coordinates $s,\phi$, defined by $p = \sqrt{s} \cos \phi, \ q = \sqrt{s} \sin \phi$, in terms of which the laplacian takes the form
\begin{equation}
    \nabla^2 = 4 \left( s \partial_s^2 + \partial_s \right) + \frac{1}{s}\partial_\phi^2,
\end{equation}
which allows us to rewrite \eqref{recorrencia1} as
\begin{equation}
\label{recorrencia2}
    o^{n+1}(s,\phi) = \left[ s - \hbar^2\left( s \partial_s^2 + \partial_s +\frac{1}{4s}\partial_\phi^2\right) \right]o^n(s,\phi)
\end{equation}
Since $\partial_\phi o(s,\phi) = 0$, and, as deduced from the recurrence relation, $\partial_\phi o^n(s,\phi) = 0 \Rightarrow \partial_\phi o^{n+1}(s,\phi) = 0$, we prove by induction that $\partial_\phi o^n(s,\phi) = 0 \ \forall \ n$, which eliminates the derivative with respect to $\phi$ from \eqref{recorrencia2}. This allows us to easily obtain the first terms in the recurrence relation, which, already expressed in terms of $p,q$, are given by
\begin{equation}
\label{wigner para potencias de p2 + q2}
    \begin{aligned}
        o^2(p,q) &= \left( p^2 + q^2 \right)^2 - \hbar^2 \\
        o^3(p,q) &= \left( p^2 + q^2 \right)^3 - 5\hbar^2\left( p^2 + q^2 \right) \\
        o^4(p,q) &= \left( p^2 + q^2 \right)^4 - 14\hbar^2\left( p^2 + q^2 \right)^2 + 5 \hbar^4 \\
    \end{aligned}
\end{equation}
We see that, in general, $\hat{o}^n(p,q)$ is a polynomial of order $n$ in $(p^2+q^2)$, whose dominant term is $(p^2+q^2)^n$, while corrections proportional to even powers of $\hbar$ are also present.

\section{Numerical Details}

The calculations in this article were performed using the Julia language \cite{Julia-2017} . The package DifferentialEquations.jl \cite{DifferentialEquations.jl-2017} was used to solve the necessary differential equations in parallel. The calculations were performed on a 12th Gen Intel Core i5-12600K processor, which has 16 threads.

\subsection{Morse System}

The integrals related to the Morse system were performed using Gaussian quadrature. The integration region, in units of $\omega = \hbar = 1$, is given by
\begin{equation}
    R = \left\{ (p,q) \in \mathbb{R}^2 \ \left| \ \chi p^2 + \frac{1}{4\chi} \left( 1-e^{-q} \right)^2 < \frac{1}{4\chi} \right. \right\}
\end{equation}
Introducing the variables $\tilde{P} = 2\chi p$ e $Q = 1-e^{-q} $, we obtain
\begin{equation}
    \begin{aligned}
        R &= \left\{ \left(\tilde{P},Q\right) \in \mathbb{R}^2 \ \left| \ \tilde{P}^2 + Q^2 < 1 \right. \right\} \\ &= \left\{ \left(\tilde{P},Q\right) \in \mathbb{R}^2 \ \left| \ Q \in \left( -1,1 \right); \  \tilde{P} \in \left( -\sqrt{1-Q^2},\sqrt{1-Q^2} \right) \right. \right\},
    \end{aligned}
\end{equation}
which can be simplified by defining $P = \tilde{P}/\sqrt{1-Q^2}$. The, we have
\begin{equation}
    R = \left\{ \left(P,Q\right) \in \mathbb{R}^2 \ \left| \ Q \in \left( -1,1 \right); \  P \in \left( -1,1 \right) \right. \right\}.
\end{equation}
The inverse transformation is then
\begin{equation}
    \begin{cases}
       p = \sqrt{1-Q^2}\dfrac{P}{2\chi} \\
       q = -\ln \left( 1-Q \right)
    \end{cases},
\end{equation}
which has jacobian determinant
\begin{equation}
    \det \frac{\partial (p,q)}{\partial (P,Q)}  = \frac{1}{2\chi} \sqrt{\frac{1+Q}{1-Q}},
\end{equation}
which is proportional to the weight function of a Gauss-Chebyshev quadrature of the 3º kind. We therefore use this quadrature rule to perform the integration over the $Q$ coordinate, while a Gauss-Legendre quadrature is used to integrate over $P$. The advantage of Gaussian quadrature is that the integration points will be independent of $\theta$, and then a single set of points can be used to compute the thermodynamic quantities over a range of temperatures. In this work, we used a grid of $300 \times 300$ points to perform the integration, which corresponds to $9 \times 10^4$ trajectories.

\subsection{Nelson System}

In the semiclassical calculations for the Nelson system, different techniques were used for different set of parameters. 

In the case of the energy, as well as the heat with $\mu = 1.5, 2$, we first performed the change of variables $(p_x,p_y,x,y) \mapsto (P_X,P_y,X,Y)$ with
\begin{equation}
    \begin{cases}
        P_x = \sqrt{\dfrac{\theta}{2}} p_x \\
        P_y = \sqrt{\dfrac{\theta}{2}} p_y \\
        X = \sqrt{\theta \mu} x \\
        Y = \sqrt{\theta} (y - x^2/2)
    \end{cases} .
\end{equation}

This transformation has unit jacobian determinant and, in terms of the new variables, we have that the classical Boltzmann's weight is simply
\begin{equation}
    e^{-\beta H} = \exp\left[ - \left( P_x^2 + P_y^2 + X^2 + Y^2 \right) \right] .
\end{equation}

The integration is then performed by an h-adaptive technique as described in \cite{genz1980remarks,berntsen1991adaptive}. The Julia implementation can be found in \cite{Integrals.jl}. We bounded the integration algorithm to use roughly $10^5$ integration points. We used the BS3 \cite{BOGACKI1989321,ODESolvers} and Vern6 \cite{verner2010,ODESolvers} algorithms to solve the differential equations, and the tolerances varied between $10^{-2}$ and $10^{-6}$. For each $\mu$, the corresponding plot took around $40$ seconds to $3$ minutes to complete.

We found that the heat capacities with $\mu = 0.5, 1$ were much harder to integrate. In this case, we didn't perform a change of variables and resorted to a Monte Carlo integration method, where the $10^7$ integration points were sampled from the \textit{classical} Boltzmann's distribution $e^{-\beta H(\boldsymbol{x})}/Z$ using the Metropolis-Hastings algorithm \cite{metropolis1953equation,hastings1970monte}. In this case, for each $\mu$, the corresponding plot took around $3$ hours to complete.

For the energy spectrum, which is used to calculate the quantum versions of the thermodynamic quantities, we used a grid of $160 \times 160$ points, where $x$ spanned from $-4.5$ to $4.5$, and $y$ spanned from $-4$ to $5$. We then approximated the laplacian of the time independent Schrödinger equation through a finite differences matrix over this grid. The discretizatation of this equation gives rise to a eigenvalue equation, which can be solve through standard linear algebra libraries.

\end{appendices}

\bibliography{sn-bibliography}% common bib file

\end{document}